\documentclass[12pt]{article}
\usepackage{amsfonts}
\usepackage{latexsym}
\usepackage{amsmath}
\usepackage{amssymb}
\usepackage{amssymb}
\usepackage{slashed}

\newcommand{\newsection}{    % Numeration of eqs. is automatic
\setcounter{equation}{0}\section}
\def\appendix#1{\addtocounter{section}{1}\setcounter{equation}{0}
\renewcommand{\thesection}{\Alph{section}}
\section*{Appendix \thesection\protect\indent \parbox[t]{11.15cm}{#1}}
\addcontentsline{toc}{section}{Appendix \thesection\ \ \ #1}}

\newcommand{\be}{\begin{eqnarray}}
\newcommand{\ee}{\end{eqnarray}}
\newcommand{\bea}{\begin{eqnarray}}
\newcommand{\eea}{\end{eqnarray}}
\newcommand{\ba}{\begin{array}}
\newcommand{\ea}{\end{array}}

\newenvironment{frcseries}{\fontfamily{frc}\selectfont}{}

\def \la {\label}

\def\a{\alpha}

\def\d{\delta}

\def\G{\Gamma}

\def\cA{{\cal A}}

\def\cL{{\cal L}}

\def\hn{{\hat{\nu}}}

\def\bbe{{\bf{e}}}
\font\mybb=msbm10 at 11pt

\def\bb#1{\hbox{\mybb#1}}

\def\bR {\bb{R}}

\def\k{\kappa}
\def\cD{{\cal D}}
\def\cI{{\cal I}}
\def\cS{{\cal S}}
\def \tn {\tilde{\nabla}}

\def\tnp{\hat{\tn}}

\def\al{{\cal{O}}(\alpha')}
\def\alsq{{\cal{O}}(\alpha'^2)}

\def\hn {{\tilde{\nabla}}}
\def\hnz {{\hn^{[0]}}{}}

\begin{document}
\begin{titlepage}
\begin{center}
%\today
\vspace*{-1.0cm}
%\hfill hep-th/yymmnnn \\
\hfill DMUS--MP--16/07 \\

\vspace{2.0cm} {\Large \bf  Anomaly Corrected Heterotic Horizons} \\[.2cm]

\vskip 2cm
A. Fontanella$^1$,~J. B.  Gutowski$^1$ and G. Papadopoulos$^2$
\\
\vskip .6cm

\begin{small}
$^1$\textit{Department of Mathematics,
University of Surrey \\
Guildford, GU2 7XH, UK. \\
Email: a.fontanella@surrey.ac.uk \\
Email: j.gutowski@surrey.ac.uk}
\end{small}\\*[.6cm]

\begin{small}
$^2$\textit{  Department of Mathematics, King's College London
\\
Strand, London WC2R 2LS, UK.\\
E-mail: george.papadopoulos@kcl.ac.uk}
\end{small}\\*[.6cm]

\end{center}

\vskip 3.5 cm
\begin{abstract}

\vskip1cm

We consider supersymmetric near-horizon geometries in heterotic supergravity
up to two loop order in sigma model perturbation theory. We identify the conditions for the horizons to admit
enhancement of supersymmetry. We show that solutions which undergo
supersymmetry enhancement exhibit an $\mathfrak{sl}(2, \bR)$ symmetry,
and we describe the geometry of their horizon sections.
We also prove a modified Lichnerowicz type theorem, incorporating
$\alpha'$ corrections,
which relates Killing spinors to zero modes of near-horizon Dirac operators.
Furthermore, we demonstrate that there
are no $AdS_2$ solutions in heterotic supergravity up to second order in $\alpha'$
for which the fields are smooth and the internal space is smooth and compact without boundary. We investigate a class
of nearly supersymmetric  horizons, for which the gravitino Killing spinor equation is satisfied on the
spatial cross sections but   not  the dilatino one,  and  present a description of their geometry.

\end{abstract}

\end{titlepage}

%%%%%%%%%%%%%%%%%%%%%%%%%%%%%%%%%%%%%%%%%%%%%%%%%%%%%%%%%%%%%%%%%%%%%%%%%%

%%%%%%%%%%%%%%%%%%%%%%%%%%%%%%%%%%%%%%%%%%%%%%%%%%%%%%%%%%%%%%%%%%%%%%%%%%
 \setcounter{section}{0}

\newsection{Introduction}

The effect of higher order corrections to supergravity solutions
is of considerable interest, perhaps most notably for our understanding of
quantum corrections to black holes. This is important in determining how string theory may resolve black hole singularities, as well as the investigation of the properties of
black holes away from the limit $\alpha' \rightarrow 0$.
In higher dimensions the four dimensional uniqueness theorems
\cite{israel, carter, hawking, robinson1, israel2, mazur, robinson}
 no longer hold, and there are exotic types of black hole
solutions, such as the five dimensional black rings \cite{Emparan:2001wn}. For
ten and eleven dimensional supergravity, it is expected that there
is a particularly rich structure of black objects, and the classification
of these is ongoing. Progress has recently been
made in the classification of the near-horizon geometries
of supersymmetric black holes. Near-horizon
geometries of extremal black holes in  supergravity are known to
generically undergo supersymmetry enhancement. This has been proven by analysing
the global properties of such solutions via generalized Lichnerowicz theorems
\cite{lichner11, lichneriib, lichneriia1, lichneriia2},
and making use of index theory arguments \cite{atiyah1}. One consequence of the supersymmetry enhancement is that
all such near-horizon geometries exhibit an $\mathfrak{sl}(2,\bR)$ symmetry.  However, it is not apparent that
these properties persist after including string theory corrections.

There are several approaches to investigate how $\alpha'$ corrections can change the event horizons of
black holes.  Many black holes have $AdS_p \times S^q$ near-horizon geometries and as it is expected that the symmetries of such backgrounds
persist in quantum theory, only the radii of the sphere and $AdS$ receive $\alpha'$ corrections. However, we
expect that exotic black holes in higher dimensions need not necessarily have
such near horizon geometries.

Another approach, in the context of supersymmetric black holes
in four and five dimensions,
is to assume that the corrected near horizon geometries
undergo an enhancement of supersymmetry in the near-horizon limit,
which simplifies considerably the analysis of the Killing spinor equations.
It is known that all supergravity  $D=4$ and $D=5$ black holes undergo supersymmetry
enhancement in the near-horizon limit \cite{Kallosh:1992ta, Ferrara:1996dd, Gibbons:1993sv}.
In particular, the five dimensional BMPV black hole \cite{Breckenridge:1996is} undergoes supersymmetry enhancement
from $N=4$ to $N=8$ (maximal supersymmetry) in the near-horizon limit \cite{Chamseddine:1996pi}.
Also, the supersymmetric asymptotically $AdS_5$ black hole
of \cite{Gutowskiads5bh} undergoes supersymmetry enhancement
from $N=2$ to $N=4$ (half-maximal supersymmetry) in the near-horizon limit.
However it is not clear in general why one expects that
the $\alpha'$ corrections should preserve this property.

The first systematic classification of supersymmetric near-horizon geometries
in a higher derivative theory in five dimensions \cite{Hanaki:2006pj} was done in
\cite{Gutowski:2011nk},
in which the only assumption made was that the solutions should preserve
the minimal amount of supersymmetry. The five dimensional theory reduces to
ungauged five-dimensional supergravity coupled to arbitrarily many vector multiplets
when the higher derivative corrections are set to zero. In this limit, it is known
that near-horizon geometries are maximally supersymmetric with constant scalars
\cite{Gutowski:2004bj}, which is consistent with the standard picture of the attractor mechanism.
In contrast, when higher derivative terms are turned on, the list of
near-horizon geometries determined in \cite{Gutowski:2011nk} includes not only
the maximally supersymmetric geometries (which were classified in \cite{Castro:2008ne}),
but also a set of regular non-maximally supersymmetric solutions, on making use
of a result of \cite{Manton:2012fv}. Although it is unclear if these particular
near-horizon geometries can be extended to a full black hole solution, the existence
of such a solution proves that for certain supergravity theories, the presence of
higher derivative terms can change how supersymmetry is enhanced for near-horizon solutions.

In this paper, we consider how higher derivative corrections to ten dimensional supergravity
affect the geometry and supersymmetry of near-horizon solutions.
We shall choose to begin this work by investigating heterotic theory which includes $\alpha'$ corrections
up to two loops in sigma model perturbation theory.
This choice is motivated by two factors. Firstly, from the perspective
of the standard supergravity, much more is known about
the geometric structure of generic supersymmetric solutions,
and near-horizon geometries. In particular, as a consequence of
the spinorial geometry classification techniques developed in
\cite{class1, class2} which were then combined with
a global analysis of near-horizon geometries in \cite{hethor},
there exists a full classification of all possible supersymmetric
near-horizon geometries in the heterotic supergravity.
Secondly, the structure of higher derivative correction terms
in the field equations, and in the Killing spinor equations,
is significantly simpler for the heterotic theory when
compared to the types of terms which arise in type II supergravity
\cite{hetpap, Callan:1991at, Howe:1992tg, gsw}, and associated references.

The method we shall use to prove our results is that first we solve the Killing spinor equations in the
near-horizon lightcone directions, and then simplify
the remaining conditions as much as possible using both
the local field equations and Bianchi identities, as well as global analysis.
For the global analysis, we shall assume that the spatial cross-section of
the event horizon is smooth and compact, without boundary,
and that all near-horizon fields are also smooth.
 As a result of this analysis, we find that there are no $AdS_2$ solutions (at zero and
first order in $\alpha'$) to heterotic supergravity,
which completes the classification of
heterotic AdS solutions in \cite{lichnerads4}.
We also show that all of the conditions of supersymmetry
reduce to a pair of  gravitino KSEs and a pair of algebraic KSEs on the spatial horizon sections.  The latter are associated
to the dilatino KSE.
Throughout, we allow for all near-horizon data, including the spinors, to receive
$\alpha'$ corrections.

Using these conditions, we show
that there is automatic supersymmetry enhancement
at both zero and first order in $\alpha'$
in the case for which there exists  negative
light-cone chirality Killing spinor $\eta_-$ up to $\alsq$ which does not vanish at zeroth order in $\alpha'$.
In this case the supersymmetry enhancement is obtained via
the same mechanism as for the  near-horizon geometries
considered in \cite{hethor} without $\alpha'$ corrections, and the solution admits an $\mathfrak{sl}(2,\bR)$
symmetry. Such horizons admit 2, 4, 6 and 8 Killing spinors and their geometry is similar to that
of horizons with vanishing anomaly contribution examined in \cite{hethor}.
The remaining case, for which the negative
light-cone chirality spinors vanish at zeroth order in $\alpha'$
remains open. We have investigated global aspects of these
solutions by considering $\alpha'$ corrections to the global
analysis carried out in \cite{hethor}, and also by constructing
generalized Lichnerowicz theorems analogous to those
proven in \cite{lichner11, lichneriib, lichneriia1, lichneriia2},
again incorporating $\alpha'$ corrections. However, in both cases,
there is an undetermined sign in the $\alsq$ terms appearing, which precludes the extension
of the maximum principle arguments to first order in $\alpha'$.

We also consider a class of near-horizon solutions which are ``nearly'' supersymmetric.  These are not supersymmetric but some of their KSEs are satisfied.  This
is motivated by the existence of WZW type of solutions to the heterotic theory with constant dilaton.  It is known  that such solutions
solve the gravitino KSE but not the dilatino one. In the present case, we consider horizons for which one of the  gravitino KSEs is satisfied\footnote{Such solutions
are not supersymmetric, and furthermore the spacetime gravitino KSE is not necessarily satisfied.} on the spatial horizon section
up order $\alsq$ but not the other and the algebraic KSEs.  After some assumptions on the form of the fields, we give a complete
description of the geometry of such solutions.

This paper is organized as follows. In section 2,  we present the fields of heterotic near horizon geometries and  we integrate up the KSEs along the lightcone directions.
In sections  3 and 4, we identify the independent KSEs  by examining the various cases that can occur
and in the process, prove that there are no $AdS_2$ solutions.
In section 5, we determine the conditions under which the horizons exhibit supersymmetry enhancement, and in section 6
we give the geometry of the horizon sections.
In section 7, we generalize the global analysis presented
near-horizon geometries in \cite{hethor} to include $\alpha'$ corrections. However because of a $\alsq$ sign ambiguity,  it is not
possible to prove that the horizon section admits a $G_2$ structure compatible with a connection with skew-symmetric torsion, as is the
case at zeroth order in $\alpha'$. We also  generalize
the Lichnerowicz type theorems to higher orders in $\alpha'$. Once again, $\alsq$ sign ambiguity means that
it is not possible to prove that zero modes of the horizon Dirac equation (at zero and
first order in $\alpha'$) satisfy the Killing spinor equations to the same order in
$\alpha'$, although the algebraic Killing spinor involving the 2-form gauge field is satisfied
to the required order in $\alpha'$.   In sections 8 and 9, we examine the geometry of nearly supersymmetric horizons focusing on those
that admit a solution to  the gravitino KSE  on the horizon spatial section, and in section 10 we give our conclusions.

The paper contains  several appendices. In appendix A, we summarize some key formulae that are used throughout
in the computations of the paper and present the field equations of the theory. In appendix B, we provide the details of part of the proof
to identify  the independent KSEs on the spatial horizon section. In section C, we present a formula which relates
the gravitino KSE to the gaugino KSE which is instrumental in the investigation of the geometry of nearly supersymmetric horizons.
In appendix D, we present further detail of the proof of the Lichnerowicz type theorem for the heterotic theory, and
in Appendix E, we describe how $AdS_{n+1}$ can be written as a warped
product over $AdS_n$, and describe how such constructions are inconsistent with
our assumptions on the global structure and regularity of the solutions.

\newsection{Supersymmetric heterotic near-horizon geometries}

\subsection{Near horizon fields}

The metric near a smooth killing horizon   expressed in Gaussian null co-ordinates
 \cite{isen, gnull} can be written as
\begin{eqnarray}
ds^2 = 2 \bbe^+ \bbe^- + \delta_{ij} \bbe^i \bbe^j~,~~~
\label{nearhormetr}
\end{eqnarray}
where  we have used the frame
\bea
\label{nhbasis}
\bbe^+ &=& du~,~~~
\bbe^- = dr + r h - {1 \over 2} r^2 \Delta du~,~~~
\bbe^i = e^i{}_J dy^J~,
\eea
$i,j=1, \dots , 8$,  $u,r$ are the lightcone coordinates, and the 1-form $h$, scalar $\Delta$
and $\bbe^i$  depend only on the  coordinates $y^I$, $I=1, \dots ,8$, transverse to the lightcone. The black hole stationary
Killing vector field is identified with $\partial_u$.
The induced metric on ${\cal S}$ is
\be
ds_{\cal{S}}^2 = \delta_{ij} \bbe^i \bbe^j
\ee
and ${\cal S}$ is taken to be compact, connected without boundary. We denote the Levi-Civita connection of ${\cal{S}}$ by $\hn$, and the Levi-Civita connection of the D=10 spacetime as
$\nabla$.

For the other heterotic fields, we assume that the dilaton $\Phi$, and the real 3-form
$H$, and non-abelian gauge potential $A$ admit well-defined near-horizon limits, and that
$\partial_u$ is a symmetry of the full solution:
\bea
{\cal{L}}_{\partial_u} \Phi=0, \qquad {\cal{L}}_{\partial_u} H = 0, \qquad {\cal{L}}_{\partial_u}A=0~.
\eea
In particular, this means that $\Phi=\Phi(y)$, and also
\bea
\label{threef}
H = \bbe^+ \wedge \bbe^- \wedge N+r \bbe^+ \wedge Y+W~,
\eea
where $N$, $Y$ and $W$ are $u,r$-independent 1, 2 and 3-forms
on ${\cal{S}}$ respectively, and we do not assume $dH=0$.
Moreover,
\bea
A= r {\cal{P}} \bbe^+ + {\cal{B}}
\eea
where ${\cal{P}}$ and ${\cal{B}}$
are a $r,u$-independent $G$-valued scalar and 1-form on ${\cal{S}}$ respectively.
The non-abelian 2-form field strength $F$ is given by
\bea
F = dA + A \wedge A~.
\eea
Our conventions for the heterotic theory including $\alpha'$ corrections
are consistent with those of \cite{hetpap}. We assume that the near-horizon data
admit a Taylor series expansion in $\alpha'$.
We denote this expansion by
\bea
\Delta = \Delta^{[0]} + \alpha' \Delta^{[1]} + \alsq
\eea
and similarly for all near-horizon data, including spinors. For the supersymmetric solutions,  we shall assume
that that there is at least one zeroth order in $\alpha'$ Killing spinor,  $\epsilon^{[0]} \neq 0$.

\subsection{Supersymmetry }

  In the previous treatments
of heterotic near-horizon geometries \cite{hethor}, it was assumed that the anomaly vanishes and
so the Bianchi identity $dH=0$ was used to further
simplify the structure of the 3-form.
Here, we shall not
take $dH=0$ as there is a non-trivial contribution from the heterotic anomaly, and so the 3-form takes the more general form
given in ({\ref{threef}}).

We remark that the KSE of
heterotic supergravity have been solved in \cite{class1}
and \cite{class2}, and so,  the solutions to the KSEs which
we consider here correspond to a subclass of the solutions
in \cite{class1, class2}. However for horizons the global assumptions on the spatial section ${\cal S}$, like compactness, allow the derivation
of additional conditions
on the spinors and on the geometry.  So it is particularly
useful to re-solve the KSEs, decomposing the spinors into
positive and negative lightcone chiralities adapted for
the Gaussian null basis (\ref{nhbasis}), $\epsilon=\epsilon_+ + \epsilon_-$, where
\bea
\Gamma_\pm \epsilon_\pm =0, \qquad \Gamma_{+-} \epsilon_\pm
= \pm \epsilon_\pm \ .
\eea
We shall then extract from the KSEs the conditions imposed on  $\epsilon_\pm$ that will be useful to apply the global conditions on ${\cal S}$.

\subsubsection{The Gravitino Equation}

We begin by considering the gravitino equation
\bea
\label{grav}
\hat\nabla_M\epsilon\equiv\nabla_M \epsilon -{1 \over 8}H_{M N_1 N_2} \Gamma^{N_1 N_2}
\epsilon= \alsq~.
\eea
First, on examining the $M=-$ component of ({\ref{grav}})
we find that
\bea
\epsilon_+ = \phi_+ + \alsq , \qquad
\epsilon_- = \phi_- + {1 \over 4} r (h-N)_i \Gamma_- \Gamma^i \phi_+ + \alsq~,
\label{grav2}
\eea
where $\partial_r \phi_\pm=0$.
Next, on examining the $M=+$ component of ({\ref{grav}}),
we find
\bea
\phi_- = \eta_- + \alsq , \qquad \phi_+ = \eta_+ + {1 \over 4}u (h+N)_i
\Gamma_+ \Gamma^i \eta_- + \alsq~,
\label{grav3}
\eea
where $\partial_r \eta_\pm = \partial_u \eta_\pm=0$.
In additon, the $M=+$ component of ({\ref{grav}}) implies a number of algebraic conditions:
\bea
\label{alg1}
\bigg({1 \over 2} \Delta +{1 \over 8}(h^2-N^2)
-{1 \over 8}(dh+Y+h \wedge N)_{ij} \Gamma^{ij} \bigg) \phi_+= \alsq~,
\eea
and
\bea
\label{alg2}
\bigg(-{1 \over 2} \Delta -{1 \over 8}(h^2-N^2)
-{1 \over 8}(dh+Y+ h \wedge N)_{ij} \Gamma^{ij} \bigg) \eta_-= \alsq~,
\eea
and
\bea
\label{alg3}
\bigg({1 \over 4} (\Delta h_i - \partial_i \Delta)\Gamma^i
-{1 \over 32} (dh+Y)_{ij}\Gamma^{ij} (h-N)_k \Gamma^k \bigg)
\phi_+= \alsq~.
\eea
We remark that ({\ref{alg1}}) and ({\ref{alg2}}) are equivalent
to
\bea
\label{alg4a}
{1 \over 2} \Delta +{1 \over 8}(h^2-N^2)= \alsq~,
\eea
\bea
\label{alg4b}
(dh+Y+ h \wedge N)_{ij} \Gamma^{ij} \phi_+= \alsq~,
\eea
and
\bea
\label{alg5b}
(dh+Y+ h \wedge N)_{ij} \Gamma^{ij} \eta_-= \alsq~,
\eea
respectively. Furthermore, using these conditions,
({\ref{alg3}}) can also be rewritten as
\bea
\label{alg6}
\bigg({1 \over 4} (\Delta h_j - \partial_j \Delta)
-{1 \over 8}(h-N)^k \big(dh+Y+2 h \wedge N)_{jk} \bigg) \Gamma^j \phi_+= \alsq~.
\eea

Next, we consider the $M=i$ components of ({\ref{grav}}).
This implies
\bea
\label{par1}
\hn_i \phi_+ + \bigg({1 \over 4}(N-h)_i -{1 \over 8} W_{ijk}
\Gamma^{jk} \bigg) \phi_+= \alsq~,
\eea
and
\bea
\label{par2}
\hn_i \eta_- + \bigg({1 \over 4}(h-N)_i -{1 \over 8} W_{ijk}
\Gamma^{jk} \bigg) \eta_-= \alsq~,
\eea
together with the algebraic condition
\bea
\label{alg7}
\bigg(\hn_i (h-N)_j + {1 \over 2}(h_i N_j - h_j N_i)
-{1 \over 2}(h_i h_j -N_i N_j)
\nonumber \\
-(dh-Y)_{ij} -{1 \over 2} W_{ijk}(h-N)^k \bigg)
\Gamma^j \phi_+= \alsq~.
\eea
These conditions exhaust the content of ({\ref{grav}}).

\subsubsection{Dilatino and Gaugino KSEs}

Next again ignoring $O(\alpha'^2)$ terms we consider the dilatino KSE
\bea
\label{akse1}
\bigg(\Gamma^M \nabla_M \Phi -{1 \over 12}H_{N_1 N_2 N_3}
\Gamma^{N_1 N_2 N_3} \bigg) \epsilon = \alsq~.
\eea
On making use of the previous conditions, it is straightforward
to show that the dilatino KSE is equivalent to
the following three conditions
\bea
\label{aksecon1}
\bigg(\Gamma^i \hn_i \Phi +{1 \over 2} N_i \Gamma^i -{1 \over 12} W_{ijk} \Gamma^{ijk} \bigg) \phi_+= \alsq~,
\eea
and
\bea
\label{aksecon2}
\bigg(\Gamma^i \hn_i \Phi -{1 \over 2} N_i \Gamma^i -{1 \over 12} W_{ijk} \Gamma^{ijk} \bigg) \eta_-= \alsq~,
\eea
and
\bea
\label{aksecon2b}
\bigg( \big(\Gamma^i \hn_i \Phi -{1 \over 2} N_i \Gamma^i -{1 \over 12} W_{ijk}
\Gamma^{ijk} \big) (h-N)_\ell \Gamma^\ell + Y_{ij} \Gamma^{ij} \bigg)
\phi_+= \alsq  \ .
\eea

It remains to consider  the gaugino KSE
\bea
\label{akse2}
F_{MN} \Gamma^{MN} \epsilon = \al~.
\eea
This implies the following conditions
\bea
\label{akseconaux1}
\bigg(2 {\cal{P}} + {\tilde{F}}_{ij} \Gamma^{ij} \bigg) \phi_+= \al~,
\eea
and
\bea
\label{akseconaux2}
\bigg(-2 {\cal{P}}+{\tilde{F}}_{ij} \Gamma^{ij} \bigg) \eta_- = \al~,
\eea
and
\bea
\label{akseconaux2b}
\bigg( {1 \over 4}\big(-2 {\cal{P}} + {\tilde{F}}_{ij} \Gamma^{ij}\big)
(h-N)_\ell \Gamma^\ell
+2\big(h {\cal{P}}+ {\cal{P}} {\cal{B}}
- {\cal{B}} {\cal{P}}-d {\cal{P}}\big)_i  \Gamma^i \bigg) \phi_+ = \al~,
\eea
where
\bea
{\tilde{F}}=d {\cal{B}} + {\cal{B}} \wedge {\cal{B}}
\eea
The conditions ({\ref{akseconaux1}}) and ({\ref{akseconaux2}}) imply that
\bea
{\cal{P}}= \al~,
\eea
and so $F={\tilde{F}} + \al$. Therefore ({\ref{akse2}}) is equivalent
to
\bea
\label{aksecon3}
{\tilde{F}}_{ij} \Gamma^{ij} \phi_+= \al~,
\eea
and
\bea
\label{aksecon4}
{\tilde{F}}_{ij} \Gamma^{ij} \eta_-=\al~,
\ee
and
\bea
\label{aksecon4b}
{\tilde{F}}_{ij} \Gamma^{ij} (h-N)_\ell \Gamma^\ell \phi_+ = \al~.
\eea

In order to simplify these conditions further,
we shall first consider the two cases for which either $\phi_+^{[0]} \equiv 0$
or  $\phi_+^{[0]} \not \equiv 0$.

\newsection{Solutions with $\phi_+^{[0]} \equiv 0$}

Suppose that there exists a Killing spinor $\epsilon$ with
$\epsilon^{[0]} \not \equiv 0$, but $\phi_+^{[0]} \equiv 0$.
Such a spinor must therefore have $\eta_-^{[0]} \not \equiv 0$, and hence
it follows that
\bea
h^{[0]}+N^{[0]}=0 \ .
\eea
Then ({\ref{par2}}) implies that
\bea
\label{partrans}
d \parallel \eta_-^{[0]} \parallel^2 = - \parallel \eta_-^{[0]} \parallel^2
h^{[0]} \ .
\eea
In particular, this condition implies that if
$\eta_-^{[0]}$ vanishes at any point on the horizon
section, then $\eta_-^{[0]}=0$ everywhere.
So, $\eta_-^{[0]}$ must be everywhere non-vanishing.

On taking the divergence of ({\ref{partrans}}), and
making use of the $N_1=+, N_2=-$ component of the 2-form gauge potential field equation ({\ref{geq1}}), one obtains the following condition

\bea
\hnz^i \hnz_i \parallel \eta_-^{[0]} \parallel^2 - \big(2 \hn^i \Phi^{[0]} + \parallel \eta_-^{[0]} \parallel^{-2} \hnz^i  \parallel \eta_-^{[0]} \parallel^2 \big) \hnz_i  \parallel \eta_-^{[0]} \parallel^2 =0 \ .
\eea
As $ \parallel \eta_-^{[0]} \parallel^2$ is nowhere vanishing, an application of the maximum principle
implies that $ \parallel \eta_-^{[0]} \parallel^2=const.$, and hence ({\ref{partrans}})
gives that
\bea
h^{[0]}=0, \qquad N^{[0]}=0 \ .
\eea
These conditions, together with  ({\ref{alg4a}}), imply that
\bea
\Delta=\alsq \ .
\eea
Then the dilaton field equation ({\ref{deq}}) implies that
\bea
\hn^i \hn_i (e^{-2 \Phi}) ={1 \over 6} e^{-2 \Phi} W_{ijk} W^{ijk} + \al~,
\eea
and hence it follows that
\bea
\Phi^{[0]}=const, \qquad W^{[0]}=0 \ .
\eea
Furthermore, this then implies that
\bea
H=du \wedge dr \wedge N +r du \wedge Y + W + \alsq~,
\eea
and hence
\bea
dH = du \wedge dr \wedge (dN-Y)-r du \wedge dY +dW + \alsq~.
\eea
As the $ruij$ component on the RHS of the Bianchi identity is $\alsq$
this implies that
\bea
Y=dN+\alsq
\eea
and in particular, $Y^{[0]}=0$.

Next consider the gauge equations. The $+-$ component of the 2-form gauge potential field equations ({\ref{geq1}}) is
\bea
\label{dfree1}
\hn^i N_i = \alsq~.
\eea
Also, the $u$-dependent part of ({\ref{par3}}) implies that
\bea
\hn_i (h+N)_j \Gamma^j \eta_- = \alsq~,
\eea
which gives that
\bea
\label{udepsimp1}
\hn_i(h+N)_j = \alsq~.
\eea
Taking the trace of this expression, and using ({\ref{dfree2}}) yields
\bea
\label{dfree2}
\hn^i h_i = \alsq~.
\eea
Next, recall that the gravitino KSE ({\ref{par4}}) implies
\bea
\label{dnsq1}
\hn_i \parallel \eta_- \parallel^2 = -{1 \over 2}(h-N)_i  \parallel \eta_- \parallel^2 + \alsq
\eea
Taking the divergence yields, together with ({\ref{dfree1}}) and ({\ref{dfree2}}) the condition
\bea
\hn^i \hn_i  \parallel \eta_- \parallel^2 = \alsq
\eea
which implies that $ \parallel \eta_- \parallel^2= const + \alsq$.
Substituting back into ({\ref{dnsq1}}) gives the condition
$N=h+\alsq$, and hence ({\ref{udepsimp1}}) implies
that
\bea
\hn_i h_j = \alsq \ .
\eea

So, to summarize, for this class of solutions, we have obtained the following
 conditions on the fields
\bea
\label{bossimp1}
N=h+\alsq, && \quad h^{[0]}=0, \quad Y=\alsq, \quad \hn_i h_j= \alsq,
\nonumber \\
\Delta = \alsq, && \quad H^{[0]}=0, \quad \Phi^{[0]}=const~,
\eea
and it is straightforward to check that the generic conditions on
$\phi_+$ then simplify to
\bea
\label{par3bb}
\hn_i \phi_+ -{1 \over 8}W_{ijk} \Gamma^{jk} \phi_+= \alsq~,
\eea
and
\bea
\label{auxalg1bbb}
\bigg(\Gamma^i \hn_i \Phi +{1 \over 2} h_i \Gamma^i -{1 \over 12} W_{ijk} \Gamma^{ijk} \bigg) \phi_+= \alsq
\eea
and
\bea
\label{auxalg1cbb}
{\tilde{F}}_{ij} \Gamma^{ij} \phi_+ = \al \ .
\eea

The generic conditions on $\eta_-$ also simplify to
\bea
\label{par4bb}
\hn_i \eta_- -{1 \over 8}W_{ijk} \Gamma^{jk} \eta_-= \alsq
\eea
and
\bea
\label{auxalg2bbb}
\bigg(\Gamma^i \hn_i \Phi -{1 \over 2} h_i \Gamma^i -{1 \over 12} W_{ijk} \Gamma^{ijk} \bigg) \eta_-= \alsq
\eea
and
\bea
\label{auxalg2cbb}
{\tilde{F}}_{ij} \Gamma^{ij} \eta_- = \al \ .
\eea

In the next section, we shall  consider the case for which there exists a Killing spinor with
$\phi_+^{[0]} \not \equiv 0$.
It will be shown that the  conditions ({\ref{bossimp1}}) on the bosonic fields
and the simplified KSEs listed above correspond to special cases
of the corresponding  conditions on the fields and simplified KSEs
of  $\phi_+^{[0]} \not \equiv 0$. In particular,
this will allow the KSEs for $\phi_+^{[0]} \equiv 0$
and $\phi_+^{[0]} \not \equiv 0$ to be written in a unified way.

\newsection{Solutions with $\phi_+^{[0]} \not \equiv 0$}

Suppose that there exists a Killing
spinor $\epsilon$, with $\epsilon^{[0]} \not \equiv 0$ and
$\phi_+^{[0]} \not \equiv 0$. Then consider ({\ref{par1}}); this implies that
\bea
\label{pt1}
\hn_i \parallel \phi_+ \parallel^2 = {1 \over 2}(h_i-N_i)\parallel \phi_+ \parallel^2 + \alsq~,
\eea
and ({\ref{alg7}}) gives that
\bea
\label{alg7b}
\hn_i (h-N)_j + {1 \over 2}(h_i N_j - h_j N_i)
-{1 \over 2}(h_i h_j -N_i N_j)
\nonumber \\
-(dh-Y)_{ij} -{1 \over 2} W_{ijk}(h-N)^k = \alsq~.
\eea
Taking the divergence of ({\ref{pt1}}), and using ({\ref{par1}})
together with the trace of ({\ref{alg7b}}), we find that
\bea
\label{lapsq1}
\hn^i \hn_i \parallel \phi_+ \parallel^2 - h^i \hn_i \parallel \phi_+ \parallel^2 = \alsq~.
\eea
An application of the maximum principle (see e.g. \cite{maxp})
then yields the condition
\bea
\hn_i \parallel \phi_+ \parallel^2= \alsq~.
\eea

To see this, note that to zeroth order in $\alpha'$,
({\ref{lapsq1}}) implies that $\hn^{[0]}_i \parallel \phi_+^{[0]}\parallel^2=0$, on applying the maximum principle.
Then ({\ref{pt1}}) and ({\ref{alg7b}}) imply that $N^{[0]}=h^{[0]}$ and $Y^{[0]}=dh^{[0]}$; and from ({\ref{alg4a}}) we also have $\Delta^{[0]}=0$.
Then it is useful to consider the  field equations of the 2-form gauge potential
({\ref{geq1}}), which imply that
\bea
\label{bcx1}
\hn^i \bigg( e^{-2 \Phi} h_i \bigg)= \al~,
\eea
and
\bea
\label{bcx2}
e^{2 \Phi} \hn^j \big(e^{-2 \Phi} dh_{ji}\big)
+{1 \over 2} W_{ijk} dh^{jk} + h^j dh_{ji}= \al~,
\eea
and the Einstein equations imply that
\bea
\label{bcx3}
{\tilde{R}}_{ij} + \hn_{(i} h_{j)} -{1 \over 4} W_{imn} W_j{}^{mn}
+2 \hn_i \hn_j \Phi  = \al~.
\eea
Using ({\ref{bcx1}}), ({\ref{bcx2}}) and ({\ref{bcx3}})
it follows that{\footnote{We remark that
the condition ({\ref{bcx4}}) was also obtained
in \cite{hethor}. In that case, a bilinear matching condition
was imposed in order to find $N^{[0]}=h^{[0]}, Y^{[0]}=dh^{[0]}$.
Here we do not assume such a bilinear matching condition, but nevertheless
we find the same condition.}}

\bea
\label{bcx4}
&&\hn^i \hn_i h^2 + (h-2 d \Phi)^j \hn_j h^2
= 2 \hn^{(i} h^{j)} \hn_{(i} h_{j)}
\cr
&&~~~~~~~~~~~~~~~~
+{1 \over 2}(dh - i_h W)_{ij} (dh-i_h W)^{ij} + \al~.
\nonumber \\
\eea
In particular, ({\ref{bcx4}}) implies that $\hn^{[0] i} h^{[0]}_i =0$
on applying the maximum principle.
It follows from ({\ref{lapsq1}}) that
\bea
\hn^{[0]i} \hn^{[0]}_i \langle \phi_+^{[0]}, \phi_+^{[1]} \rangle
- h^{[0]i}  \hn^{[0]}_i \langle \phi_+^{[0]}, \phi_+^{[1]} \rangle =0~.
\eea
On multiplying this condition by $\langle \phi_+^{[0]}, \phi_+^{[1]} \rangle$
and integrating by parts, using $\hn^{[0] i} h^{[0]}_i =0$, one finds that
$\hn^{[0]}_i \langle \phi_+^{[0]}, \phi_+^{[1]} \rangle =0$ as well.
So, it follows that $\hn_i \parallel \phi_+ \parallel^2 = \alsq$.

Then, ({\ref{pt1}}) also implies that $N=h+\alsq$.
Substituting these conditions back into ({\ref{alg4a}}),
we find that $\Delta^{[1]}=0$ as well, so $\Delta=\alsq$.
Also, ({\ref{alg7b}}) implies that
\bea
Y -dh= {\cal{O}}(\alpha'^2)~.
\eea

To summarize the conditions on the bosonic fields;
we have shown that for solutions with $\phi_+^{[0]} \neq 0$, we must have
\bea
\label{bossimp2}
\Delta=\alsq, \qquad N=h+\alsq, \qquad Y=dh+\alsq
\eea
which implies that
\bea
H = d (\bbe^- \wedge \bbe^+) + W + \alsq~.
\eea
The  field equation ({\ref{geq1}}) of the 2-form gauge potential
can then be rewritten in terms of the near-horizon data
as
\bea
\label{geq1a}
\hn^i \big( e^{-2 \Phi} h_i \big)= \alsq~,
\eea
\bea
\label{geq1b}
e^{2 \Phi} \hn^j \big(e^{-2 \Phi} dh_{ji}\big)
+{1 \over 2} W_{ijk} dh^{jk} + h^j dh_{ji}= \alsq~,
\eea
and
\bea
\label{geq1c}
e^{2 \Phi} \hn^k \big(e^{-2 \Phi} W_{kij}\big)
+ dh_{ij} - h^k W_{kij} = \alsq~.
\eea
In addition, ${\cal{P}}=\al $ and so $F= {\tilde{F}} + \al$.
The $i,j$ component of the Einstein equation then simplifies to
\bea
\label{einsp}
{\tilde{R}}_{ij} + \hn_{(i} h_{j)} -{1 \over 4} W_{imn} W_j{}^{mn}
+2 \hn_i \hn_j \Phi
\nonumber \\
+{\alpha' \over 4} \bigg(-2 dh_{i \ell}
dh_j{}^\ell + \check {\tilde{R}}_{i \ell_1, \ell_2 \ell_3}
\check {\tilde{R}}_j{}^{\ell_1, \ell_2 \ell_3}
- {\tilde{F}}_{i\ell}{}^{ab} {\tilde{F}}_j{}^\ell{}_{ab} \bigg) =\alsq~.
\eea
Furthermore, dilaton field equation can be written as
\bea
\label{deqsimp1}
\hn^i \hn_i \Phi - h^i \hn_i \Phi -2 \hn^i \Phi \hn_i \Phi -{1 \over 2} h_i h^i
+{1 \over 12} W_{ijk} W^{ijk}
\nonumber \\
+{\alpha' \over 16} \big(2 dh_{ij} dh^{ij}
+ {\tilde{F}}_{ij}{}^{ab} {\tilde{F}}^{ij}{}_{ab}
- \check {\tilde{R}}_{\ell_1 \ell_2, \ell_3 \ell_4}
\check {\tilde{R}}^{\ell_1 \ell_2, \ell_3 \ell_4} \big) = \alsq \ .
\eea

On making use of the conditions (\ref{bossimp2})  on the bosonic fields, the KSEs on
$\phi_+$ then simplify further to
\bea
\label{par3}
\hn_i \phi_+ -{1 \over 8}W_{ijk} \Gamma^{jk} \phi_+= \alsq~,
\eea
\bea
\label{auxalg1}
dh_{ij} \Gamma^{ij} \phi_+= \alsq~,
\eea
\bea
\label{auxalg1b}
\bigg(\Gamma^i \hn_i \Phi +{1 \over 2} h_i \Gamma^i -{1 \over 12} W_{ijk} \Gamma^{ijk} \bigg) \phi_+= \alsq~,
\eea
and
\bea
\label{auxalg1c}
{\tilde{F}}_{ij} \Gamma^{ij} \phi_+ = \al \ .
\eea

Furthermore, KSEs on $\eta_-$ also simplify to
\bea
\label{par4}
\hn_i \eta_- -{1 \over 8}W_{ijk} \Gamma^{jk} \eta_-= \alsq~,
\eea
\bea
\label{auxalg2}
dh_{ij} \Gamma^{ij} \eta_-= \alsq~,
\eea
\bea
\label{auxalg2b}
\bigg(\Gamma^i \hn_i \Phi -{1 \over 2} h_i \Gamma^i -{1 \over 12} W_{ijk} \Gamma^{ijk} \bigg) \eta_-= \alsq~,
\eea
and
\bea
\label{auxalg2c}
{\tilde{F}}_{ij} \Gamma^{ij} \eta_- = \al \ .
\eea
In both cases above, (\ref{par3}) and (\ref{par4}) are a consequence of the gravitino KSE, (\ref{auxalg1b}) and (\ref{auxalg2b}) are associated to the dilatino KSE,
while (\ref{auxalg1c}) and (\ref{auxalg2c}) are derived from the gaugino KSE. The two additional conditions (\ref{auxalg1}) and (\ref{auxalg2}) can be thought of
as integrability conditions.

\subsection{Independent KSEs}

The KSEs we have stated in the previous sections (\ref{par3bb})-(\ref{auxalg2cbb}) and (\ref{par3})-(\ref{auxalg2c}) are not all independent.
It turns out that  the independent KSEs  are
\bea
\label{gravsimp}
\tnp\eta_\pm\equiv \hn_i \eta_\pm - {1 \over 8} W_{ijk} \Gamma^{jk} \eta_\pm = \alsq
\eea
and
\bea
\label{algsimpmax}
\bigg(\Gamma^i \hn_i \Phi \pm {1 \over 2} h_i \Gamma^i -{1 \over 12} W_{ijk} \Gamma^{ijk} \bigg) \eta_\pm = \alsq \ .
\eea
This is the case irrespectively on whether $\phi_+^{[0]} \equiv 0$ or $\phi_+^{[0]} \not= 0$  though the
conditions on the bosonic fields are somewhat different. The proof of this independence of the KSEs requires the use of field equations and Bianchi identities
and it is rather involved.  The details  can be found in appendix B.

\newsection{Supersymmetry enhancement}

A key ingredient in the investigation of heterotic horizons is that supersymmetry always enhances.  As a result horizons preserve
2, 4, 6 and 8 supersymmetries \cite{hethor}.  However this is based on a global argument which we shall see does not necessarily
apply to $\alsq$.

As a result we shall seek some alternative conditions to guarantee that supersymmetry enhances. In particular we shall show that
if there exists a Killing spinor $\epsilon=\epsilon(\eta_+, \eta_-)$ up to  $\alsq$, ie $\eta_-$
solves (\ref{gravsimp}) and (\ref{algsimpmax}) up to  $\alsq$,
such that $\eta_-^{[0]} \neq 0$, and the horizon has $h^{[0]} \neq 0$, then  there is automatic supersymmetry enhancement.

To prove this, it suffices to demonstrate that $h$ leaves all fields invariant and that it is covariantly constant with respect
to the connection with torsion $\tnp$ on ${\cal S}$.  Indeed, first note that ({\ref{udepa}}) implies that
\bea
\label{niceh}
\tnp_ih_j\equiv \hn_i h_j - {1 \over 2} W_{ijk} h^k= \alsq~.
\eea
In particular, to both zeroth and first order in $\alpha'$,
$h$ defines an isometry on ${\cal{S}}$, with $h^2=const+\alsq$.
Then the gauge equation ({\ref{geq1a}})
implies
\bea
\label{phlie}
{\cal{L}}_h \Phi = \alsq~.
\eea
Also, the $u$-dependent part of ({\ref{auxalg1c}}) implies
\bea
\label{extraalg3}
(i_h {\tilde{F}})_i \Gamma^i \eta_-= \al~,
\eea
which implies that $i_h {\tilde{F}}= \al$. So
in the gauge for which $i_h {\cal{B}}=0$, one has
\bea
{\cal{L}}_h {\tilde{F}} = \al \ .
\eea
Next we consider ${\cal{L}}_h W$, where
\bea
\label{lie3}
{\cal{L}}_h W = -{\alpha' \over 2}   \bigg( {\rm tr}\big( (i_h \check R) \wedge \check R\big) \bigg)+\alsq~,
\eea
because $dh=i_h W+\alsq$. To evaluate this expression, note first that the integrability conditions of
\bea
\tnp_i \eta_-=\alsq, \qquad \tnp_i(h_\ell \Gamma^\ell \eta_-)=\alsq
\eea
are
\bea
{\hat {\tilde{R}}}_{ijpq} \Gamma^{pq} \eta_-=\alsq, \qquad
{\hat {\tilde{R}}}_{ijpq} \Gamma^{pq} (h_\ell \Gamma^\ell \eta_-)=\alsq
\eea
from which we obtain the condition
\bea
h^\ell {\hat {\tilde{R}}}_{ij\ell q} =\alsq~,
\eea
and hence, as a consequence of ({\ref{curvcross}}),
\bea
h^\ell \check{\tilde{R}}_{\ell qij} =\al~.
\eea
Moreover,
\bea
h^\ell  \check {\tilde{R}}_{\ell q+-} = h^i (dh)_{i q} =\alsq~.
\eea
It follows that the contribution of $i_h \check R$ to the RHS of ({\ref{lie3}}) is of at least $\al$, and hence
\bea
{\cal{L}}_h W=\alsq \ .
\eea
So, we have shown that to both zero and first order in $\alpha'$,
the Lie derivative of the metric on ${\cal{S}}$, as well as $h, \Phi$ and $W$ with respect to $h$
vanishes, and the Lie derivative of ${\tilde{F}}$ with respect to $h$ vanishes to zeroth order
in $\alpha'$.

Supersymmetry is therefore enhanced, because  if $\eta_+$ satisfies ({\ref{gravsimp}})
and ({\ref{algsimpmax}}), then so does $\eta_-' = \Gamma_- h_i \Gamma^i \eta_+$.  Conversely, if $\eta_-$ satisfies
({\ref{gravsimp}})
and ({\ref{algsimpmax}}), then so does $\eta_+'= \Gamma_+ h_i \Gamma^i \eta_-$.
The proof of this makes use of the conditions
({\ref{niceh}}), together with ({\ref{phlie}}) and ({\ref{auxalg1}})
and ({\ref{auxalg2}}), and the reasoning is identical to that used
in \cite{hethor}.
This establishes  a 1-1 correspondence between
spinors $\eta_+$ and $\eta_-$ satisfying ({\ref{gravsimp}})
and ({\ref{algsimpmax}}), so the number of supersymmetries preserved
is always even.

Next  we wish to
determine whether a similar supersymmetry enhancement argument   holds for $\eta_+$ spinors. In particular  if there exists a solution to ({\ref{gravsimp}})
and ({\ref{algsimpmax}}) with $\eta_+^{[0]} \neq 0$ and  $h^{[0]} \neq 0$, does this
imply that the number of $\eta_+$ solutions is equal to the number of $\eta_-$ solutions?
This does not follow
from a local analysis of ({\ref{gravsimp}})
and ({\ref{algsimpmax}}), because there is no analogue of
the condition ({\ref{udepa}}) acting on $\eta_+$.
Nevertheless, in \cite{hethor} a global analysis was used in
order to establish such a correspondence, by computing the
Laplacian of $h^2$ and applying a maximum principle argument,
in order to obtain ({\ref{niceh}}) to zeroth order in $\alpha'$.
We shall revisit this analysis in section \ref{hsq} including
the $\alpha'$ corrections.

\newsection{Geometry}

It is a consequence of the results of \cite{hethor}, see also section \ref{hsq}, that
horizons with non-trivial fluxes preserve an even number of supersymmetries  up to ${\cal O}(\alpha')$. Furthermore we have also demonstrated that such horizons
with $\eta_-$ Killing spinors preserve an even number of supersymmetries up to ${\cal O}(\alpha'^2)$. It is straightforward to see
that horizons with more than 8 supersymmetries are trivial, ie the rotation $h$ vanishes. Therefore, the heterotic horizons of interest preserve
2,4,6 and 8 supersymmetries.

Up to ${\cal O}(\alpha')$, the investigation of geometry of all such horizons is identical to that given in \cite{hethor} for heterotic horizons
with closed 3-form field strength. Here we shall describe the geometry of the horizons that admit a  $\eta_-$ Killing spinor up to ${\cal O}(\alpha'^2)$. We have seen that
 for such horizons $h$ is parallel with respect to the connection with torsion up to ${\cal O}(\alpha'^2)$. Because of this,   the geometry of such horizons is very similar to that
 of horizons with closed 3-form flux. The only differences between the geometries of the two cases  are solely located in the modified Bianchi identity for the 3-form flux.
 As the two cases are similar, the description of the geometry will be brief.

%As we have pointed out if $h$ non-zero, the horizons at ${\cal O}(\alpha')$ admit an even number of supersymmetries. The failure of horizons to  admit an even number of
%supersymmetries at order ${\cal O}(\alpha'^2)$ can only be due to the failure of $\eta_-$ spinors to solve the KSEs up to ${\cal O}(\alpha'^2)$. This is rather unusual.
%This is because it would mean that there is an anomaly in preserving spacetime supersymmetries by higher order $\alpha'$ terms.  Equivalently, it would mean that there
%is an anomaly for $h$ to be parallel with respect to the connection with torsion at higher orders in $\alpha'$. So it is likely that such a case does not arise but
%we shall not explore this here.

\subsection{Horizons with $G_2$ structure}

Such horizons admit two supersymmetries up to ${\cal O}(\alpha'^2)$.  In particular $h$ satisfies (\ref{niceh}).
The spacetime locally can be described as a (principal)  $SL(2, \bR)$ fibration over a 7-dimensional manifold $B^7$
which admits a metric $d\tilde s_{(7)}^2$ and a 3-form $\tilde H_{(7)}$ such that the connection $\hat{\tilde\nabla}^{(7)}$  with torsion   $\tilde H_{(7)}$
has holonomy contained in $G_2$.
The spacetime metric and 3-form flux can be written as
\bea
ds^2&=&\eta_{ab} \lambda^a \lambda^b+d\tilde s_{(7)}^2+{\cal O}(\alpha'^2)~,~~~
\cr
H&=&CS(\lambda)+\tilde H_{(7)}+{\cal O}(\alpha'^2)~,
\eea
where $CS(\lambda)$ is the Chern-Simons form\footnote{Note that $CS(\lambda)= du\wedge dr\wedge h+r du\wedge dh+k^{-2} h\wedge dh$.} of the principal bundle connection,
\bea
\lambda^- &=& \bbe^-~,~~~
\lambda^+ = \bbe^+ - {1 \over 2} k^2 u^2 \bbe^- -u h~,~~~
\lambda^1 = k^{-1} \big(h+ k^2 u \bbe^-\big)~,
\la{g2vbi}
\eea
$k^2=h^2$ is constant up to ${\cal O}(\alpha'^2)$
and
\bea
\tilde H_{(7)}=k  \varphi+ e^{2\Phi} \star_7d\big( e^{-2\Phi} \varphi\big)+\alsq~.
\eea
The 3-form $\varphi$ is the fundamental $G_2$ and it is related to the fundamental $Spin(7)$ form of the $\eta_+$ Killing spinor via $\varphi=k^{-1} i_h\phi+\alsq$.
The associated vector fields to $\lambda^-, \lambda^+, \lambda^1$ satisfy a $\mathfrak{sl}(2,\bR)$ algebra. The dilaton $\Phi$ depends only on the coordinates of $B^7$.

To find solutions, one has to solve the remaining equations
\bea
&&d[e^{-2\Phi}\star_7\varphi]={\cal O}(\alpha'^2)~,~~~
\cr
&&k^{-2}\,dh\wedge dh+ d\tilde H_{(7)}=-{\alpha'\over4} \bigg(-2 dh\wedge dh+ \mathrm {tr}( \check R_{(8)}\wedge \check R_{(8)}- F\wedge F)\bigg)+{\cal O}(\alpha'^2)~,~~~
\cr
&&(dh)_{ij}={1\over2} \star_7\varphi_{ij}{}^{kl}
(dh)_{kl}+{\cal O}(\alpha'^2)~,~~~~F_{ij}={1\over2} \star_7\varphi_{ij}{}^{kl}
F_{kl}+{\cal O}(\alpha'^2)~.
\label{g2cons}
\eea
The first condition in (\ref{g2cons}) is required for $B^7$ to admit a $G_2$ structure compatible with a connection with skew-symmetric torsion. The  second condition
is the anomalous Bianchi identity of the 3-form field strength written in terms of $B^7$ data. The curvature $\check R_{(8)}$ is that of the near horizon section ${\cal S}$
with metric and skew symmetric torsion given by
\bea
d\tilde s_{(8)}^2= k^{-2} h\otimes h+d\tilde s_{(7)}^2+\alsq~,~~~\tilde H_{(8)}= k^{-2}  h\wedge dh+\tilde H_{(7)}+\alsq~.
\eea
As $\check R_{(8)}$ is invariant under $h$ and $i_h \check R_{(8)}=\alsq$, it descends on $B^7$. Finally, the last two equations in (\ref{g2cons})
imply that both $dh$ and $F$ are $\mathfrak{g}_2$ instantons on $B^7$.

\subsection{Horizons with $SU(3)$ structure}

Such horizons preserve 4 supersymmetries.  Locally the spacetime is a principal bundle with fibre $SL(2, \bR)\times U(1)$ over a K\"ahler with torsion manifold (KT) $B^6$
with Hermitian form $\omega_{(6)}$.
The metric and 3-form field strength of the spacetime can be written as
\bea
ds^2=\eta_{ab} \lambda^a \lambda^b+ d\tilde s^2_{(6)}+{\cal O}(\alpha'^2)~,~~~H&=&CS(\lambda)+\tilde H_{(6)}+{\cal O}(\alpha'^2)~,
\eea
where $\lambda^a$, $a=+,-,1,6$ is the principal bundle connections whose $a=+,-,1$ components are as in (\ref{g2vbi}) and
\bea
\lambda^6=k^{-1} \ell
\eea
which is along the $\mathfrak{u}(1)$ direction in the Lie algebra. $h^2=k^2$ is constant up to ${\cal O}(\alpha'^2)$. The curvature of the principal bundle connection
$\lambda^a$ is expressed in terms of $dh$ and $d\ell$ which are 2-forms on $B^6$ and it is required to satisfy that
\bea
dh^{2,0}=d\ell^{2,0}=\alsq~,~~~dh_{ij} \omega_{(6)}^{ij}=\alsq~,~~~d\ell_{ij} \omega_{(6)}^{ij}=-2 k^2+\alsq~,
\eea
ie $h$ is a $\mathfrak{su}(3)$ instanton on $B^6$ while $\ell$ is a $\mathfrak{u}(3)$ instanton on $B^6$.

The KT manifold $B^6$ is in addition conformally balanced, ie
\bea
\theta_{\omega_{(6)}}=2d\Phi+\alsq~,
\eea
where $\theta$ is the Lee form and the torsion is
\bea
\tilde H_{(6)}=-i_I d\omega+\alsq =e^{2 \Phi} \star_6 d [e^{-2\Phi} \omega_{(6)}]+\alsq~.
\eea
The dilaton $\Phi$ depends only on the coordinates of $B^6$.  The gauge connection is a $\mathfrak{su}(3)$ instanton on $B^6$, i.e.
\bea
F^{2,0}=\al~,~~~F_{ij} \omega_{(6)}^{ij}=\al~.
\eea

To find examples for such horizons two additional conditions should be satisfied.  One is the restriction that
\bea
\hat{\tilde R}_{(6)}{}_{ij} \omega_{(6)}^{ij}=-2 k^2 d\ell+\alsq~.
\eea
This arises from requirement that the $U(3)$ structure on $B^6$ lifts to a $SU(3)$ structure on the spacetime or equivalent the spatial horizon section ${\cal S}$. The other is
the anomalous Bianchi identity which now reads
\bea
&&k^{-2} dh\wedge dh+k^{-2} d\ell\wedge d\ell+ d\Big(e^{2 \Phi}\star_6 d [e^{-2\Phi} \omega]\Big)=
\cr~~~~~~~~~~~&&-{\alpha'\over4} \bigg(-2 dh\wedge dh+ \mathrm {tr}( \check R_{(8)}\wedge \check R_{(8)}- F\wedge F)\bigg)+{\cal O}(\alpha'^2)~,
\eea
where $\check R_{(8)}$ is the curvature of the connection with torsion on ${\cal S}$ which now its metric and torsion are given by
\bea
d\tilde s^2&=&k^{-2} (h\otimes h+\ell\otimes\ell)+d\tilde s_{(6)}^2+\alsq~,
\nonumber \\
\tilde H&=& k^{-2}  (h\wedge dh+\ell\wedge d\ell)+\tilde H_{(6)}+\alsq~.
\eea
Note that $\hat\nabla_{(8)}$ has holonomy contained in $SU(3)$ and so $\check R_{(8)}$ is a well defined form on $B^6$.
\subsection{Horizons with $SU(2)$ structure and 6 supersymmetries}

The spacetime is locally a $SL(2,\bR)\times SU(2)$ principal fibration over a 4-dimensional anti-self-dual Weyl Einstein manifold $B^4$
with metric $d\mathring s^2_{(4)}$ and quaternionic K\"ahler structure 2-forms $\omega^{r'}_{(4)}$. The spacetime metric and 3-form field strength can be expressed as
\bea
ds^2=\eta_{ab} \lambda^a\lambda^b+ \delta_{r's'} \lambda^{r'} \lambda^{s'}+e^{2\Phi} d\mathring s^2_{(4)}+\alsq~,~~~H=CS(\lambda)+\tilde H_{(4)}+\alsq~,
\eea
where $\tilde H_{(4)}=-\mathring\star_{4} de^{2\Phi}$, the principal bundle connection $\lambda^a$ for $a=+,-,1$ coincides with that
of (\ref{g2vbi}) while
\bea
\lambda^{r'}=k^{-1} \ell^{r'}~,
\eea
are the components along the $\mathfrak{su}(2)$ subalgebra of the fibre.  Furthermore the dilaton depends only on the coordinates of $B^4$, $dh$ as well as the
curvature $({\cal F}^{\rm sd})^{r'}$ of $\lambda^{r'}$ are 2-forms on $B^4$. In addition, we have that
\bea
dh^{\rm sd}=\alsq~, ~~~({\cal F}^{\rm sd})^{r'}={k\over4}\omega_{(4)}^{r'}+\alsq~,~~~F^{\rm sd}=\al
\la{6conx}
\eea
and $dh^{\rm ad}$,   $({\cal F}^{\rm ad})^{r'}$ and $F^{\rm ad}$  are not restricted, where the self-dual and anti-self dual components are appropriately denoted.
Geometrically, the set up is such that the $SO(4)=SU(2)\cdot SU(2)$ structure of $B^4$ when lifted the 7-dimensional manifold  which is the principal
bundle with fibre $SU(2)$ reduces to $SU(2)$ as required from supersymmetry.

The only remaining condition to find solutions is
\bea
 &&\mathring{\nabla}^2 e^{2\Phi}=-{1\over2} ({\cal F}^{\rm ad})_{ij}^{r'}({\cal F}^{\rm ad})^{ij}_{r'}-{k^{-2}\over2}
 dh_{ij} dh^{ij}+{3\over 8} k^2 e^{4\Phi}
 \cr&&~~~~~~~~~~~~~~~+{\alpha'\over8} \bigg(-2  dh_{ij} dh^{ij}+ \mathrm {tr}( \check R_{(8) ij} \check R_{(8)}{}^{ij}- F_{ij} F^{ij})\bigg)+{\cal O}(\alpha'^2)~.
 \label{6horcon}
\eea
Again $\check R_{(8)}$ is the curvature of the connection with torsion of the horizon section ${\cal S}$ which has metric and 3-form field strength
\bea
d\tilde s^2&=&k^{-2} h\otimes h+\delta_{r's'} \lambda^{r'} \lambda^{s'}+ e^{2\Phi} d\mathring s^2_{(4)}+\alsq~,
\nonumber \\
\tilde H &=& k^{-2}  h\wedge dh+CS(\lambda^{r'})+\tilde H_{(4)}+\alsq~.
\eea
As $\hat R_{(8)}$ has holonomy contained in $SU(2)$, $\check R_{(8)}$ is a 2-form on $B^4$. For more details on the geometry of heterotic backgrounds
that preserve 6 supersymmetries and have $SU(2)$ holonomy see \cite{compgp, hethor}.

\subsection{Horizons with $SU(2)$ structure and 8 supersymmetries}

This class of horizons have a similar geometry to those of the previous section that preserve 6 supersymmetries.  The differences are that
\bea
({\cal F}^{\rm sd})^{r'}=\alsq~,
\eea
so ${\cal F}^{r'}$ is an anti-self dual instanton on $B^4$ which now is a hyper-K\"ahler manifold with respect to the metric $d\mathring s^2_{(4)}$.  Furthermore
the equation for the dilaton (\ref{6horcon}) now reads
\bea
 &&\mathring{\nabla}^2 e^{2\Phi}=-{1\over2} {\cal F}_{ij}^{r'}{\cal F}^{ij}_{r'}-{k^{-2}\over2}
 dh_{ij} dh^{ij}
 \cr&&~~~~~~~~~~~+{\alpha'\over8} \bigg(-2  dh_{ij} dh^{ij}+ \mathrm {tr}( \check R_{(8) ij} \check R_{(8)}{}^{ij}- F_{ij} F^{ij})\bigg)+{\cal O}(\alpha'^2)~.
 \label{8horcon}
\eea
Therefore at zeroth order, a partial integration argument reveals that
\bea
dh={\cal O}(\alpha')~,~~~{\cal F}^{r'}={\cal O}(\alpha')~.
\eea
Thus $B^4$  up to a local isometry is $AdS_3\times S^3\times T^4$ or $AdS_3\times S^3\times K_3$ and the dilaton is constant. One does not expect additional $\alpha'$ corrections
to the geometry in the case that the $\check R_{(8)}$ is identified with $F$.  Though additional corrections are expected otherwise. In the absence of 5-branes, consistency
requires that the Pontryagin number of the tangent bundle of $B^4$ cancels that of the gauge bundle which is the vanishing condition for the global anomaly.

\newsection{Global Properties}

\subsection{ Maximum principle on  $h^2$} \label{hsq}

We shall revisit the global analysis of
\cite{hethor} by calculating the Laplacian of $h^2$,
but including also $\alpha'$ correction terms. Then we shall
examine the conditions imposed on the geometry by
this expression. To avoid the trivial case when $h^2=\alsq$, we take
$h^{[0]} \neq 0$.

Next we calculate the Laplacian of $h^2$ to find
that
\bea
\label{lap1}
&&\hn^i \hn_i h^2 + (h-2 d \Phi)^j \hn_j h^2
= 2 \hn^{(i} h^{j)} \hn_{(i} h_{j)}
+{1 \over 2}(dh - i_h W)_{ij} (dh-i_h W)^{ij}
\nonumber \\
&&~~-{\alpha' \over 4} h^i h^j
\bigg(-2 dh_{i \ell}
dh_j{}^\ell + \check {\tilde{R}}_{i \ell_1 \ell_2 \ell_3}
\check {\tilde{R}}_j{}^{\ell_1 \ell_2 \ell_3}
- {\tilde{F}}_{i\ell}{}^{ab} {\tilde{F}}_j{}^\ell{}_{ab} \bigg) + \alsq~.
\eea
In computing this expression, we
 made use of the Einstein equation ({\ref{einsp}})
together with the gauge field equations ({\ref{geq1a}}) and
({\ref{geq1b}}).
We remark that the calculation proceeds
in exactly the same way as in \cite{hethor}; the $\alpha'$ terms
in ({\ref{lap1}}) originate from the $\alpha'$ terms in
$2 h^i h^j {\tilde{R}}_{ij}$.
It should be noted
that in order to fully control $\alsq$ terms in this expression,
one would require to know the Einstein equations up to and including
$\alpha'^2$.

To begin, we consider ({\ref{lap1}}) to zeroth order in $\alpha'$.
We then re-obtain the conditions found in \cite{hethor} via a maximum
principle argument, i.e.
\bea
\label{firstiso}
h^2&=& {\rm const} + \al~,~~~
\hn_{(i} h_{j)}=  \al~,~~~
dh-i_h W = \al
\eea
In particular, it follows from these conditions that
\bea
i_h dh=O(\alpha') \ ,
\eea
and also
\bea
{\cal{L}}_h \Phi = \al, \qquad {\cal{L}}_h W = \al \ .
\eea
Furthermore, it also follows that if $\eta_+$ satisfies
({\ref{gravsimp}}), then $\Gamma_- h_i \Gamma^i \eta_+$ also satisfies
({\ref{gravsimp}}) to zeroth order in $\alpha'$. The integrability conditions
therefore imply that
\bea
{\hat{\tilde{R}}}_{ijmn} h^m \Gamma^n \phi_+ = \al~,
\eea
and hence
\bea
\check {\tilde{R}}_{mnij} h^m = \al~.
\eea

On substituting these conditions back into ({\ref{lap1}}) one finds that
the remaining content of ({\ref{lap1}}) is
\bea
\label{lap2}
\hn^i \bigg( e^{-2 \Phi} \hn_i h^2 \bigg) + e^{-2 \Phi} h^j \hn_j h^2
= {\alpha' \over 2} e^{-2 \Phi} h^i h^j {\tilde{F}}_{i\ell}{}^{ab} {\tilde{F}}_j{}^\ell{}_{ab}
+ \alsq~.
\eea
On integrating the $\al$ part of ({\ref{lap2}}) over the zeroth order
horizon section, one finds that
\bea
i_h {\tilde{F}} = \al~,
\eea
and furthermore
\bea
h^2 = {\rm const} + \alsq~.
\eea
It should be noted however that ({\ref{lap1}}) does not in general imply ({\ref{niceh}}). In particular, the conditions obtained from
the analysis of the properties of $h^2$ are not sufficient to
imply that if $\eta_+$, with $\eta_+^{[0]} \neq 0$, satisfies
({\ref{gravsimp}})
and ({\ref{algsimpmax}}), then $\eta_-'' =  \Gamma_- h_i \Gamma^i
\eta_+$ also satisfies ({\ref{gravsimp}})
and ({\ref{algsimpmax}}). Thus although ({\ref{lap1}}) implies the horizons exhibit supersymmetry enhancement at $\al$, it   does not imply
the same at  $\alsq$.

\subsection{ Lichnerowicz Type Theorems}

Next we shall investigate whether it is possible to identify Killing spinors with the zero modes of a suitable Dirac-like operator, by constructing a
generalized Lichnerowicz type theorem which incorporates the near-horizon fluxes.
Such Lichnerowicz type theorems have been established for near-horizon geometries
in D=11 supergravity \cite{lichner11}, type IIB \cite{lichneriib} and type IIA supergravity (both massive and massless) \cite{lichneriia1, lichneriia2},
as well as for $AdS$ geometries in ten and eleven dimensional supergravity \cite{lichnerads1, lichnerads2, lichnerads3, lichnerads4}.

To begin, let us first define the modified connection with torsion and the modified horizon Dirac operator, respectively
\begin{align}
{\nabla}^{(\kappa)}_{i} \equiv \tnp_{i} + \k \, \G_{i} \cA  \ , \qquad\qquad\quad
\cD  \equiv \G^{i} \tnp_{i} + q \, \cA \ ,
\end{align}
where $\k, q \in \mathbb{R}$, and
\begin{align}
\notag
\tnp_i \eta_{\pm} &= \tn_i \eta_{\pm} - \frac{1}{8} W_{ijk}\G^{jk} \eta_{\pm}  \ , \\
\cA &= W_{ijk}\G^{ijk} - 12\G^i \tn_i \Phi \mp 6 \G^i h_i \ .
\end{align}
It is clear that if $\eta_{\pm}$ is a Killing spinor, i.e.
\bea
\tnp_i \eta_{\pm} = \alsq, \qquad {\rm and} \qquad \cA \eta_\pm = \alsq \ ,
\eea
then $\cD \eta_{\pm} = \alsq$ also. Here we want to investigate the extent to
which the converse is true. We shall show that if $\cD \eta_{\pm} = \alsq$, then
\bea
\label{killsp1}
\tnp_i \eta_{\pm} = \al, \qquad {\rm and} \qquad \cA \eta_\pm = \al \ ,
\eea
and moreover
\bea
\label{killsp2}
{{dh}}_{ij} \Gamma^{ij} \eta_\pm = \al, \qquad {\rm and} \qquad {{\tilde{F}}}^{ab}_{ij}\Gamma^{ij} \eta_\pm
= \al \ .
\eea

In order to obtain this result, we begin by considering the following functional
\be
\label{I functional}
\cI \equiv  \int_{\cS} e^{c\Phi} \bigg( \langle {\nabla}^{(\kappa)}_{i} \eta_{\pm} , {\nabla}^{(\kappa)i} \eta_{\pm} \rangle
-  \langle\cD \eta_{\pm} , \cD \eta_{\pm} \rangle \bigg) \ ,
\ee
where $c \in \mathbb{R}$, and we assume all the  field equations. After some algebra, which is described in appendix D, we find
\begin{align}
\label{final_I}
\notag
\cI  = &\left(8\k^2 - \frac{1}{6} \k \right) \int_{\cS} e^{-2 \Phi} \parallel \cA\,  \eta_{\pm} \parallel^2
+ \int_{\cS} e^{-2\Phi} \langle \eta_{\pm}, \Psi \cD \eta_{\pm} \rangle \\
&- \frac{\a'}{64} \int_{\cS} e^{-2\Phi} \left( 2 \parallel \slashed{dh}\, \eta_{\pm} \parallel^2 + \parallel \slashed{\tilde{F}} \eta_{\pm} \parallel^2 - \langle \check{\tilde{R}}_{\ell_1\ell_2,\, ij}\G^{\ell_1\ell_2}\eta_{\pm}, \check{\tilde{R}}^{ ij}_{\ell_3\ell_4,}\G^{\ell_3\ell_4}\eta_{\pm}\rangle \right) + \alsq\ ,
\end{align}
which is true if and only if $q= \frac{1}{12} + \alsq$ and $c = -2 +\alsq$, and the  $\Psi$ is defined as follows
\be
\Psi \equiv 2\left(\k - \frac{1}{12}\right) \cA^{\dagger} -2 \G^{i}\tilde{\nabla}_{i} \Phi - \frac{1}{6} \G^{\ell_1\ell_2\ell_3}W_{\ell_1\ell_2\ell_3} + \alsq \ .
\ee
The values of $q$ and $c$ are fixed by requiring that certain terms in the
functional ({\ref{I functional}}), which cannot be rewritten in terms of
the Dirac operator ${\cal{D}}$, or ${\cal{A}}^\dagger {\cal{A}}$, and which have no fixed sign, should vanish.

The part of (\ref{final_I}) which is of zeroth order in $\a'$ implies that if $0 < \k < \frac{1}{48}$, then
\be
\label{Dirac->Killing}
\label{gravitino + alg}
\cD \eta_{\pm} = \alsq  \quad \Longrightarrow  ({\ref{killsp1}})
\ee
and establishes the first part of the theorem. Next
the integrability condition of $\tnp\eta_{\pm} = \al$ is
\be
\hat{\tilde{R}}_{mn, \ell_1\ell_2}\G^{\ell_1\ell_2}\eta_{\pm} = \al \ ,
\ee
which in turn implies that
\be
\check{\tilde{R}}_{\ell_1\ell_2, mn}\G^{\ell_1\ell_2} \eta_{\pm} = \al \ .
\ee
Hence we shall neglect the term in (\ref{final_I}) which is quadratic in
$\check{\tilde{R}}$, as this term is ${\cal{O}}(\alpha'^3)$.
Then, assuming (\ref{Dirac->Killing}), the part of (\ref{final_I})
which is first order in $\alpha'$ further implies ({\ref{killsp2}}).
This completes the proof.

\newsection{Nearly supersymmetric horizons }

\subsection{Description of the backgrounds}

We have proven that for near horizon geometries the necessary and sufficient conditions
imposed by supersymmetry on the spinors can be reduced
to ({\ref{gravsimp}}) and ({\ref{algsimpmax}}).
In this section, we shall consider
the case for which the
supersymmetry is explicitly partially broken, in the sense that the
gravitino KSE ({\ref{gravsimp}})
admits solutions but not  dilatino one  ({\ref{algsimpmax}}). We also assume that the  fields satisfy
\bea
\Delta = \alsq, \qquad H = d (\bbe^- \wedge \bbe^+) +W + \alsq \ .
\label{nearh}
\eea
These conditions were previously obtained via the supersymmetry
analysis; here we shall assume them.
In particular, all of the conditions obtained from
the global analysis of the Laplacian of $h^2$ in Section 7
remain true. As a consequence of this,
\bea
\label{zeroiso}
\tnp_i h_j = \al~.
\eea
However we do not assume that $\tnp h = \alsq$.

One consequence of these assumptions is that none of the spacetime Killing spinor equations are satisfied even at $\al$.
In particular, the spacetime gravitino KSE requires in addition the condition that $dh_{ij}\Gamma^{ij}\eta_+=\al$ which is
not one of our requirements. In what follows, we shall investigate the consequences of the above assumptions on the geometry of
the spatial horizon sections ${\cal S}$.  We shall also comment on the special case where $\tnp h = \alsq$.

\subsection{Additional parallel spinors}

A key property of backgrounds that satisfy the gravitino KSE but not the dilatino one is the existence of
additional parallel spinors, see also appendix C.  In the present context to show this focus
 on the spinor $\eta_+$;
a similar analysis can be undertaken for the $\eta_-$ spinors.
To proceed, it will be useful to define
\bea
{\cA} = W_{ijk} \Gamma^{ijk} -12 \Gamma^i \hn_i \Phi -6 h_i \Gamma^i~,
\eea
so that the algebraic condition ({\ref{algsimpmax}}) on $\eta_+$ is equivalent to ${\cal{A}} \eta_+  = \alsq$.
We then note the useful identity
\bea
\hn_i W_{\ell_1 \ell_2 \ell_3} \Gamma^{\ell_1 \ell_2 \ell_3} \eta_+ &=&
\hn_i (\cA \eta_+) -{1 \over 8} W_{i \ell_1 \ell_2} \Gamma^{\ell_1 \ell_2}
(\cA \eta_+)
\nonumber \\
&+&3 W_{\ell_1 \ell_2 q} W_{i \ell_3}{}^q \Gamma^{\ell_1 \ell_2 \ell_3} \eta_+
-\big(6 \hn^m \Phi+3 h^m\big) W_{mi\ell} \Gamma^\ell \eta_+
\nonumber \\
&+&\big(12 \Gamma^\ell \hn_i \hn_\ell \Phi +6 \hn_i h_\ell \Gamma^\ell\big) \eta_+ \ .
\eea
The integrability conditions of ({\ref{gravsimp}}) imply that
\bea
\label{ksenil1}
{1 \over 6} \bigg(\hn_i (\cA \eta_+) -{1 \over 8} W_{i \ell_1 \ell_2}
\Gamma^{\ell_1 \ell_2} (\cA \eta_+) \bigg)
-{\alpha' \over 8} ({\tilde{F}}_{i \ell})_{ab} \Gamma^\ell ({\tilde{F}}_{q_1 q_2})^{ab}
\Gamma^{q_1 q_2} \eta_+
\nonumber \\
-{\alpha' \over 16}  dh_{i \ell} \Gamma^\ell dh_{q_1 q_2} \Gamma^{q_1 q_2} \eta_+ = \alsq~,
\eea
and hence
\bea
{1 \over 6} \langle \eta_+, \Gamma^i \hn_i(\cA \eta_+) -{1 \over 8} W_{\ell_1 \ell_2 \ell_3}
\Gamma^{\ell_1 \ell_2 \ell_3} (\cA \eta_+) \rangle
+{\alpha' \over 8} \langle (({\tilde{F}}_{\ell_1 \ell_2})_{ab} \Gamma^{\ell_1 \ell_2} \eta_+
, ({\tilde{F}}_{q_1 q_2})^{ab} \Gamma^{q_1 q_2} \eta_+ \rangle
\nonumber \\
+{\alpha' \over 16} \langle dh_{\ell_1 \ell_2} \Gamma^{\ell_1 \ell_2} \phi_+,
dh_{q_1 q_2} \Gamma^{q_1 q_2} \eta_+ \rangle = \alsq \ .
\nonumber \\
\eea
Integrating this expression over ${\cal{S}}$ yields the conditions
\bea
\label{F_dh_cond}
{\tilde{F}}_{ij} \Gamma^{ij} \eta_+ = \al, \qquad
dh_{ij} \Gamma^{ij} \eta_+ = \al~,
\eea
and substituting these conditions back into ({\ref{ksenil1}}) then implies
that
\bea
\hn_i (\cA \eta_+) -{1 \over 8} W_{i \ell_1 \ell_2} \Gamma^{\ell_1 \ell_2} (\cA \eta_+)=\alsq \ .
\label{naeta}
\eea
Therefore the spinor $\tau_+=\cA \eta_+$ is also $\hat{\tilde\nabla}$-parallel. As $\tau_+$ has opposite chirality from $\eta_+$ cannot be identified as an additional Killing spinor within the heterotic theory. Nevertheless it is instrumental in the description of the geometry of ${\cal S}$.

\subsection{Nearly supersymmetric horizons with $G_2$ holonomy}

\subsubsection{A symmetry of horizon section}

Suppose that we consider solutions
for which there exists a single solution $\eta_+$ to
the gravitino KSE
\bea
\label{covcon1}
\tnp \eta_+ = \alsq~,
\eea
for which $\big({\cal{A}} \eta_+\big)^{[0]} \neq 0$.
This implies that the   horizon section
${\cal{S}}^{[0]}$ at zeroth order in $\alpha'$ admits a $G_2$ structure.

We begin by defining $\tau_+ = {\cal{A}} \eta_+$, with $\tau_+^{[0]} \neq 0$. It will be particularly useful to define
\bea
\label{vecV}
V_i = \langle \eta_, \Gamma_i \tau_+ \rangle~.
\eea
In what  follows we shall show that $V$ is a symmetry of all the fields of the spatial
horizon section.

As $\tau_+^{[0]} \neq 0$, this  implies that $V^{[0]} \neq 0$.
In addition, as $\eta_+$ and $\tau_+$ satisfy
\bea
\label{eta_tau}
\tnp \eta_+ = \alsq~, \qquad \tnp \tau_+
=\alsq~,
\eea
it follows that
\bea
\label{parallelV}
\tnp V = \alsq~,
\eea
so that $V^2=const. + \alsq$, and $V$ is an isometry of
${\cal{S}}$ to both zero and first order in $\alpha'$.

Next, we consider the relationship of $V$ to $h$. In particular, the
spinors $h_i \Gamma^i {\cal{A}} \eta_+$ and $V_i \Gamma^i {\cal{A}} \eta_+$
are both parallel with respect to $\tnp$ at zeroth order in $\alpha'$. As we have assumed that ({\ref{covcon1}}) admits only one
solution, there must be a nonzero constant $c$ such that
\bea
V = ch+ \al~.
\eea
In addition, we have
\bea
\cL_V W = i_V dW + \alsq~,
\eea
because $dV=i_V W + \alsq$. Also, as $V=ch+\al$ it follows that
\bea
\cL_V W = c i_h dW + \alsq~.
\eea
As a consequence of ({\ref{zeroiso}}), one has  that $i_h dh=\al$,
and from the global analysis of the Laplacian of $h^2$, we find
$i_h {\tilde{F}}=\al$ as well as  $\check{\tilde{R}}_{mnij} h^m = \al$.
These conditions imply that
\bea
\cL_V W = \alsq~,
\eea
and so $W$ is invariant.

Next we consider $\cL_V \Phi$. As $V = c h + \al$ it follows that
\bea
\cL_V dh = c \cL_h dh + \al = \al~.
\eea
Also we have
\bea
\cL_V {\tilde{R}}_{ij,pq} = \alsq~,
\eea
and
\bea
\big(\cL_V {\tilde{F}}\big)_{ij}{}^a{}_b \tilde{F}^{ijb}{}_a
= \al~,
\eea
which follows from
\bea
\cL_V {\tilde{F}} = c [{\tilde{F}}, i_h {\cal{B}}]+ \al~.
\eea
Hence we have
\bea
\cL_V \bigg( \alpha' \big(-2dh_{ij} dh^{ij} + \check{\tilde{R}}_{ij,pq}
\check{\tilde{R}}^{ij,pq} - ({\tilde{F}}_{ij})^{ab} ({\tilde{F}}^{ij})_{ab} \big) \bigg) = \alsq~.
\eea
So, on taking the Lie derivative of the trace of ({\ref{einsp}}) with respect to $V$ we find
\bea
\label{L_trEins}
\cL_V \bigg( \hn^i h_i +2 \hn_i \hn^i \Phi \bigg) =\alsq~,
\eea
and hence, as a consequence of the field equation ({\ref{geq1a}}), we find
\bea
\label{liex1}
\cL_V \bigg( h^i \hn_i \Phi + \hn^i \hn_i \Phi \bigg) = \alsq~.
\eea
Also, on taking the Lie derivative of the dilaton field equation
({\ref{deqsimp1}}), we get
\bea
\label{liex2}
\cL_V \bigg(-h^i \hn_i \Phi -2 \hn_i \Phi \hn^i \Phi + \hn^i \hn_i \Phi \bigg) = \alsq
\eea
On taking the sum of ({\ref{liex1}}) and ({\ref{liex2}}), we find
\bea
\cL_V \bigg( \hn^i \hn_i \Phi - \hn^i \Phi \hn_i \Phi \bigg) = \alsq
\eea
and hence if $f= \cL_V \Phi$ we have
\bea
\label{laplx3}
\hn_i \hn^i f -2 \hn^i \Phi \hn_i f = \alsq~.
\eea
We know $\cL_h \Phi = \al$ as a consequence of the analysis
of the Laplacian of $h^2$, so $f=\alpha' f^{[1]}+ \alsq$.
Then, on integrating, ({\ref{laplx3}}) implies that
\bea
\int_{{\cal{S}}^{[0]}} e^{-2 \Phi^{[0]}} \hn_i f^{[1]} \hn^i f^{[1]} = 0~,
\eea
so $f^{[1]}=\beta$ for constant $\beta$, and so
\bea
\cL_V \Phi = \beta \alpha' + \alsq~.
\eea
As we require that $\Phi$ must attain a global maximum on ${\cal{S}}$,
at this point $\cL_V \Phi=0$ to all orders in $\alpha'$, for any $V$.
This fixes $\beta=0$, so
\bea
\label{L_Phi}
\cL_V \Phi = \alsq~,
\eea
which proves the invariance of $\Phi$.

Next, we consider $\cL_V h$. On taking the Lie derivative of the field  equation of the 2-form
gauge potential ({\ref{geq1c}}) we find
\bea
\label{Lie_gauge}
d (\cL_V h)_{ij} - (\cL_V h)^k W_{ijk} = \alsq~,
\eea
and on taking the Lie derivative of the Einstein equation
({\ref{einsp}}) we get
\bea
\label{Lie_einst}
\hn_{(i} (\cL_V h)_{j)} = \alsq~,
\eea
where we have used
\bea
\cL_h \bigg( {\tilde{F}}_{i\ell}{}^{ab} {\tilde{F}}_j{}^\ell{}_{ab} \bigg) = \al~.
\eea
It follows that
\bea
\tnp (\cL_V h)_j = \alsq~.
\eea
As $V=ch+ \al$, it is convenient to write
\bea
\cL_V h = \alpha' \Lambda+\alsq~,
\eea
where
\bea
\tnp \Lambda = \al~.
\eea
As $\Lambda_j \Gamma^j {\cal{A}} \eta_+$ and $h_j \Gamma^j {\cal{A}} \eta_+$
are both parallel with respect to $\tnp$ at zeroth order in $\alpha'$,
it follows as a consequence of (ii) that we must have
\bea
\Lambda = b h + \al~,
\eea
for constant $b$. It is also useful to compute
\bea
\label{hL_h}
h^i (\cL_V h_i) = h^i \bigg( V^j \hn_j h_i + h_j \hn_i V^j \bigg)
= {1 \over 2} \cL_V h^2 + h^i h^j \hn_i V_j = \alsq~,
\eea
which follows because $h^2 = \mathrm{const} + \alsq$, and $\tnp V = \alsq$.
This implies  that $b=0$, and hence
\bea
\cL_V h = \alsq~.
\eea
So $V$ is a symmetry of the full solution to both zeroth and first order in
$\alpha'$.

\subsubsection{Geometry}

We have shown that $V$ is a symmetry of the backgrounds up ${\cal O}(\alpha'^2)$. To investigate further the geometry
of the horizon section ${\cal S}$, let us first consider the consequences of the existence of the $\eta_+$ Killing spinor.
As the isotropy group of $\eta_+$ in $Spin(8)$ is $Spin(7)$, the fundamental self-dual 4-form $\phi$ of $Spin(7)$ on ${\cal S}$ is $\hat{\tilde \nabla}$-parallel.
It is known that in such a case, the torsion 3-form $W$ can be uniquely determined in terms of $\phi$ and the metric without any additional
conditions on the $Spin(7)$ structure of ${\cal S}$ \cite{ivanovspin7}. Next the condition  $\hat{\tilde \nabla}\tau_+=\alsq$ with  $\tau_+={\cal A} \eta_+$ is equivalent
to requiring that
\bea
\hat{\tilde \nabla}_i\big(( 2d\Phi+h)_j-(\theta_\phi)_j\big)={\cal O}(\alpha'^2)~,
\eea
where $\theta_\phi$ is the Lee form of $\phi$, see \cite{class1}.  As a result $2d\Phi+h-\theta_\phi$ is a parallel 1-form.  If it is not linearly dependent on $V$, it will give rise
to an additional solution for the gravitino KSE ${\cal S}$.  As we have assumed that there is strictly one parallel spinor of the same chirality as $\eta_+$, we have to require that
\bea
2d\Phi+h-\theta_\phi=\lambda V +\alsq~,
\label{phivi}
\eea
for some non-zero constant $\lambda$; for $\lambda=0$ the dilatino KSE is satisfied as well.

Let us next turn to investigate the $G_2$ structure on ${\cal S}$.  As $V$ is an isometry on ${\cal S}$ and $i_VW=dV$, setting $V^2=\ell^2+\alsq$ for $\ell$ constant, we can decompose
the metric and 3-form as
\bea
d\tilde s^2={1\over\ell^2} V\otimes V+ds^2_{(7)}+\alsq~,~~~W=\ell^{-2}V\wedge dV+W_{(7)}+\alsq~,
\eea
where $ds^2_{(7)}$ is the metric on the space orthogonal to $V$ and $i_VW_{(7)}=0$. The data $(ds^2_{(7)}, W_{(7)})$ are thought (locally) as the metric
torsion on the space of orbits $M^7$ of $V$.  For this observe that ${\cal L}_V W_{(7)}=0$ and as $i_V W_{(7)}=0$, $W_{(7)}$  descends as a 3-form on the space of orbits.

The spatial horizon section ${\cal S}$ admits a $G_2$ structure with fundamental form $\varphi={\ell}^{-1} i_V\phi$ as $\hat{\tilde \nabla} \varphi=\alsq$. The question is
whether this $G_2$ structure descends on the space of orbits of $V$. First observe that $i_V\varphi=0$.  So it remains to investigate whether ${\cal L}_V\varphi=\alsq$.
For this notice that under $G_2$ representations $dV$ decomposes as $dV=dV^{\bf 7}+ dV^{\bf 14}+\alsq$ because $i_V dV=\alsq$. Then use (\ref{bianx}) together with
$\hat{\tilde \nabla} \varphi=\hat{\tilde \nabla} V=\alsq$ and $i_V dW={\cal O}(\alpha'^2)$ to show that
\bea
\hat{\tilde \nabla} dV^{\bf 7}=\alsq~.
\eea
As $ dV^{\bf 7}$ is a vector in ${\cal S}$ orthogonal to $V$, if it is not vanishing will generate an additional $\hat{\tilde \nabla}$-parallel spinor  on ${\cal S}$ of the same chirality as $\eta_+$.  As we have restricted
the number of such  spinors to one, we have to set $dV^{\bf 7}=\alsq$. It has been shown in \cite{class1} that a $\hat{\tilde \nabla}$-parallel k-form $\alpha$ is invariant under the action of a
 $\hat{\tilde \nabla}$-parallel vector  $V$,
iff the rotation $i_VW$ leaves the form invariant.  As $i_V W=dV+\alsq$ and $dV$ takes values in $\mathfrak{g}_2$, we conclude that
\bea
{\cal L}_V \varphi=\alsq~.
\eea
and so $M^7$ admits a $G_2$ structure compatible with connection with skew-symmetric torsion given by the data $(ds^2_{(7)}, W_{(7)})$.  In such a
case $W_{(7)}$ can be determined uniquely in terms of $\varphi$ and $ds^2_{(7)}$ provided
a certain geometric constraint is satisfied \cite{ivanovg2}.

It remains to explore (\ref{phivi}) from the perspective of $M^7$. Let us decompose $h=V+ h^\perp$, where $g(V, h^\perp)=0$.  Then (\ref{phivi}) can be written as
\bea
&&\ell^{-1} g(V, h)-{1\over 6} ( W_{(7)})_{ijk} \varphi^{ijk}=\lambda\ell +\alsq~,
\cr
&& 2d\Phi+h^\perp-\theta_\varphi=\alsq~,
\eea
where $\theta_\varphi$ is the Lee form of $\varphi$ on $M^7$. The former determines the singlet part of $ W_{(7)}$ in terms of $V$ and $h$ while
the latter imposes the dilatino KSE on $M^7$.

\newsection{Nearly supersymmetric horizons with additional parallel spinors}

\subsection{Nearly supersymmetric horizons with $SU(3)$ holonomy}

\subsubsection{Symmetries of horizon section}\label{symsu3}

Suppose there are exactly two
linearly independent spinors $\eta^{(1)}_+$, $\eta^{(2)}_+$ such that
\bea
\label{grav_SU(3)}
\tnp \eta^{(a)}_+ = \alsq, \qquad a=1,2 \ ,
\eea
for which $\big({\cal{A}} \eta^{(a)}_+\big)^{[0]} \neq 0$, ($a=1, 2$).
It follows that the
 horizon section ${\cal{S}}^{[0]}$ admits a $SU(3)$ structure at zeroth order in $\alpha'$.

We set   $\tau^{(a)}_+ = \cA \eta^{(a)}_+$ which are non-vanishing spinors that satisfy
\bea
\label{gravtau}
\tnp \tau^{(a)}_+ = \alsq~, \qquad a=1,2 \ .
\eea
Using these we  define the 1-form and 2-form spinor bilinears $V$ and $\omega$
by
\bea
V_i = \langle \eta^{(1)}_+ , \G_i \tau^{(1)}_+ \rangle~, \qquad
\omega_{ij} = \langle \eta^{(1)}_+ , \G_{ij} \eta^{(2)}_+ \rangle~,
\eea
and also let
\bea
\tilde{V} = i_{V} \omega~.
\eea
Observe that
\bea
\label{par_cond}
\tnp V = \alsq~, \qquad \tnp \omega = \alsq~, \qquad \tnp \tilde{V} = \alsq \ .
\eea
We also define ${\tilde{h}}$ by
\be
\tilde{h} = i_h \omega \ ,
\ee
which satisfies
\bea
\label{covdd1}
\tnp \tilde{h}= \al \ .
\eea
The main task below is to show that both $V$ and $\tilde V$  leave invariant all the fields
on ${\cal S}$, and that they generate a $\bR\oplus \bR$ lie algebra.

As $V$ and $\tilde V$ are $\tnp$-parallel, they are killing. Next consider the invariance
of $W$.
The spinors $V^j \G_j \cA \eta^{(a)}_+$, $h^j \G_j \cA \eta^{(a)}_+$ and $\tilde{h}^j \G_j \cA \eta^{(a)}_+$ are all parallel with respect to $\tnp$ to zeroth order in $\alpha'$. In order for ({\ref{grav_SU(3)}})
to have exactly two solutions, we must have
\be
V = ch + \tilde{c} \tilde{h} + \al \ ,
\ee
for some constants $c$, $\tilde{c}$. Thus
\be
\cL_V W = c i_h dW + \tilde{c} i_{\tilde{h}} dW + \alsq \ .
\ee
To continue,
since the two spinors $\eta^{(1)}_+$ and $\eta^{(2)}_+$ must satisfy (\ref{F_dh_cond}), it follows that, at zeroth order in $\alpha'$, $\tilde{F}$ and $dh$ are $(1,1)$ traceless with respect to
the almost complex structure obtained from $\omega$.
This, together with the conditions $i_h dh = \al$ and $i_h \tilde{F} = \al$,
which follow from the global analysis of the Laplacian of $h^2$, implies that
\be
\label{ht_dh_F}
i_{\tilde{h}} dh = \al \ , \qquad\qquad i_{\tilde{h}} \tilde{F} = \al \ ,
\ee
and hence
\bea
i_V dh = \al \ , \qquad \qquad i_V \tilde{F}= \al \ .
\eea
It is also useful to consider the spinors $\eta^{(a)}_{+}$ and $\tilde{h}_{\ell}\G^{\ell}\eta^{(a)}_{+}$. The integrability conditions of
\be
\tnp\eta^{(a)}_{+} = \alsq \ , \qquad\qquad \tnp\left(\tilde{h}_{\ell}\G^{\ell}\eta^{(a)}_{+}\right) = \al \ ,
\ee
are
\be
\hat {\tilde{R}}_{ij, pq} \G^{pq} \eta^{(a)}_{+} = \alsq \ , \qquad\qquad
\hat {\tilde{R}}_{ij, pq} \G^{pq} \left(\tilde{h}_{\ell}\G^{\ell}\eta^{(a)}_{+}\right) = \al \ ,
\ee
which imply
\be
\label{ht_R}
\tilde{h}^p \check{\tilde{R}}_{pq, ij} = \al \ .
\ee
It follows that $i_h dW=\alsq$ and $i_{\tilde{h}} dW=\alsq$, as
a consequence of the Bianchi identity, and therefore $i_V dW= \alsq$.
Thus we have shown that
\be
\cL_V W = \alsq \ .
\ee
This proves the invariance of $W$.

Next we consider $\cL_V \Phi$. It follows from ({\ref{covdd1}}) that
\be
i_{\tilde{h}} d\tilde{h} = \al \ ,
\ee
and also
\be
\cL_{\tilde{h}} W = \al \ .
\ee
Since $\tilde{h}$ is an isometry of $\cS$ to zeroth order in $\alpha'$, we also have
\be
\cL_{\tilde{h}} \tilde{R}_{ij, pq} = \al \ .
\ee
On taking the Lie derivative of the trace of (\ref{einsp}) with respect to $\tilde{h}$, we find
\be
\cL_{\tilde{h}} \left( \tn_i \tn^i \Phi \right) = \al \ ,
\ee
which is equivalent, if $g = \cL_{\tilde{h}} \Phi$, to
\be
\label{lap_g}
\tn_i \tn^i g = \al \ .
\ee
On integrating the zeroth order of (\ref{lap_g}), we find
\be
\int_{\cS^{[0]}} \tn_i g^{[0]} \tn^i g^{[0]} = 0 \ ,
\ee
so $g^{[0]} = \gamma$, for constant $\gamma$.  Thus
\be
\cL_{\tilde{h}} \Phi = \gamma + \al \ .
\ee
Since $\Phi$ must attain a global maximum on $\cS$, at this point $\cL_{\tilde{h}} \Phi = 0$ to all orders in $\alpha'$. This fixes the constant $\gamma = 0$, and so
\be
\label{L_ht}
\cL_{\tilde{h}} \Phi = \al \ ,
\ee
which implies
\be
\label{L_Phi_al}
\cL_V \Phi = \al \ .
\ee
As $V= ch + \tilde{c}\tilde{h} + \al$, it follows that
\be
\label{L_dh}
\cL_V dh = c\cL_h dh + \tilde{c}\cL_{\tilde{h}} dh + \al = \al \ .
\ee
Since $V$ is an isometry of $\cS$ to first order in $\alpha'$, we have
\be
\label{L_R}
\cL_V \tilde{R}_{ij, pq} = \alsq \ .
\ee
Also we have
\be
\label{L_F}
(\cL_V\tilde{F})_{ij}{}^a{}_b \tilde{F}^{ij\, b}{}_a = \al \ ,
\ee
which follows from
\be
\cL_V \tilde{F} = c [{\tilde{F}}, i_h {\cal{B}}] + \tilde{c} [{\tilde{F}}, i_{\tilde{h}} {\cal{B}}] + \al
\ .
\ee
Using the conditions (\ref{L_Phi_al}), (\ref{L_dh}), (\ref{L_R}) and (\ref{L_F}), we follow the analysis for the $G_2$ case of the previous section  undertaken from the equation (\ref{L_trEins}) to (\ref{L_Phi}), and conclude that
\be
\cL_V \Phi = \alsq \ ,
\ee
which proves the invariance of the dilaton $\Phi$.

Next we consider $\cL_V h$. Equations (\ref{Lie_gauge}) and (\ref{Lie_einst}), which have been established  in the previous section, hold here as well after using in the addition  that
\be
\cL_{\tilde{h}}\bigg( \tilde{F}_{i\ell}{}^{ab}\tilde{F}_j{}^{\ell}{}_{ab} \bigg) = \al \ .
\ee
Then it follows that
\be
\tnp_i \left(\cL_V h \right)_j = \alsq \ .
\ee
Furthermore we notice that
\be
\cL_{\tilde{h}} h = \al \ .
\ee
As $V = ch + \tilde{c}\tilde{h} + \al$, it is convenient to write
\be
\cL_V h = \alpha' \Psi + \alsq \ ,
\ee
where
\be
\tnp \Psi = \al \ .
\ee
Then it follows that the spinors $\Psi_j \G^j \cA \eta_+$,  $h_j \G^j \cA \eta_+$ and $\tilde{h}_j \G^j \cA \eta_+$ are all parallel with respect to $\tnp$ at zeroth order in $\alpha'$. In order for
({\ref{grav_SU(3)}}) to admit exactly two solutions, we must have
\be
\Psi = b h + \tilde{b}\tilde{h} + \al \ ,
\ee
for constants $b$ and $\tilde{b}$. Then using  $i_h\cL_V h = \alsq$, which has been computed in (\ref{hL_h}), and  $h^2 = const. + \alsq$, it follows that $b = \al$ and therefore
\be
\cL_V h = \alpha' \tilde{b} \tilde{h} + \alsq \ .
\ee

Next we consider the symmetries generated by $\tilde{V}$. Since $V = ch + \tilde{c}\tilde{h} + \al$, then we have
\be
\tilde{V} = c\tilde{h} - \tilde{c} h + \al \ .
\ee
Since $V$ and $\omega$ are both parallel with respect to $\tnp$ to first order in $\alpha'$, we also have
\be
\tnp \tilde{V} = \alsq \ .
\ee
Then the analysis undertaken for $V$ holds as well for $\tilde{V}$, because the only properties of $V$ used through the analysis are that $V$, at zeroth order in $\alpha'$, is a linear combination of $h$ and $\tilde{h}$ with constant coefficients, and $V$ is parallel with respect to $\tnp$ to first order in $\alpha'$. Thus we argue in a similar way that
\be
\cL_{\tilde{V}} W = \alsq \ , \qquad \cL_{\tilde{V}} \Phi = \alsq \ , \qquad \cL_{\tilde{V}} h = \alpha' \tilde{q} \tilde{h} + \alsq\ ,
\ee
for a constant $\tilde{q}$.

%Finally we remark that $V$ and $\tilde{V}$ are two isometries of $\cS$ which commute to zeroth order in $\alpha'$, i.e.
%\be
%\cL_V \tilde{V} = c^2 \cL_h \tilde{h} - \tilde{c}^2 \cL_{\tilde{h}} h + \al = \al \ .
%\ee

Finally, the $V$ and $\tilde V$ commute up to $\alsq$.  To see this observe that since $i_V \tilde V=0$ and $i_V W=dV+\alsq$,  we have that
\bea
{\cal L}_{\tilde V} V=i_{\tilde V} i_V W+\alsq~.
\eea
Using (\ref{bianx}) adapted to ${\cal S}$ as well as $i_V dW=i_{\tilde V} dW=\alsq$, we conclude that
\bea
\hat{\tilde \nabla} i_{\tilde V} i_V W=\alsq~.
\eea
Therefore the vector $i_{\tilde V} i_V W$ is $\hat{\tilde \nabla}$-parallel and moreover is orthogonal to both $V$ and $\tilde V$. So if it is non-zero, it will generate
additional $\tnp$-parallel $\eta_+$ spinors on ${\cal S}$.  As we have restricted those to be strictly two, we conclude that $i_{\tilde V} i_V W$ vanishes and so
\bea
[V, \tilde V]=\alsq~.
\eea
In particular as $i_V \tilde V=0$, we have that
\bea
i_V d\tilde V=i_{\tilde V} dV=\alsq~.
\label{vdtv}
\eea
This concludes the examination of symmetries of ${\cal S}$.
\subsubsection{Geometry}

It is clear from the examination of the symmetries of the fields on ${\cal S}$ and in particular (\ref{par_cond}) and (\ref{vdtv}) that we can set
\bea
d\tilde s^2&=&\ell^{-2} V\otimes V+ \ell^{-2} \tilde V\otimes \tilde V+ ds^2_{(6)}+\alsq~
\nonumber \\
W&=&\ell^{-2} V\wedge dV+\ell^{-2} \tilde V \wedge d\tilde V+ W_{(6)}+\alsq~,
\eea
where $V^2=\tilde V^2=\ell^2+\alsq$ and $\ell$ is constant, $ds^2_{(6)}$ is the metric in the orthogonal complement of $V$ and $\tilde V$ and $i_VW_{(6)}=i_{\tilde V} W_{(6)}=\alsq$.

From construction ${\cal S}$ admits an $SU(3)$ structure.  We shall now investigate whether this (locally) descends on the space of orbits $M^6$ of $V$ and $\tilde V$.
First the data $(ds^2_{(6)}, W_{(6)})$ define a Riemannian geometry on $M^6$  with skew-symmetric torsion. In particular for the torsion this follows from
$i_VW_{(6)}=i_{\tilde V} W_{(6)}=\alsq$ and  ${\cal L}_V W_{(6)}= {\cal L}_{\tilde V} W_{(6)}=\alsq$.

Next consider the reduction of the (almost) Hermitian form $\omega$. Choosing without loss of generality $V$ and $\tilde V$ orthogonal, one can write
\bea
\omega=\ell^{-2} V \wedge \tilde V+\omega_{(6)}+\alsq~,
\eea
where $i_V\omega_{(6)}=i_{\tilde V}\omega_{(6)}=\alsq$.  For $\omega_{(6)}$ to descend to a Hermitian structure on $M^6$, it must be invariant under the action of both
$V$ and $\tilde V$.  Observe that $\hat{\tilde \nabla} \omega_{(6)}=\alsq$ and also $\hat{\tilde \nabla} V=\hat{\tilde \nabla} \tilde V=\alsq$. Thus $\omega_{(6)}$ is invariant
iff the rotations $i_V W=dV+\alsq$ and $i_{\tilde V} W=d{\tilde V}+\alsq$ leave $\omega_{(6)}$ invariant \cite{class1}. In turn this implies that the (2,0) and (0,2) parts of
the rotations which we denote with $[dV]^{2,0}$ and $[d\tilde V]^{2,0}$, respectively, must vanish. Using (\ref{bianx}), $\hat{\tilde \nabla}\omega_{(6)}=\alsq$ and $i_VdW=i_{\tilde V} dW=\alsq$, we find
that
\bea
\hat{\tilde \nabla}[i_{ V} W]^{2,0}=\hat{\tilde \nabla}[i_{\tilde V} W]^{2,0}=\alsq~.
\eea
As ${\cal S}$ has an $SU(3)$ structure compatible with $\hat{\tilde \nabla}$, contracting with the (3,0)-form both $[i_{\tilde V} W]^{2,0}$ and $[i_{\tilde V} W]^{2,0}$
give rise to vector fields in ${\cal S}$ orthogonal to both $V$ and $\tilde V$ which are $\hat{\tilde \nabla}$-parallel.  Thus the requirement of strictly
two $\eta_+$ $\tnp$-parallel  spinors leads to setting $[i_{\tilde V} W]^{2,0}=[i_{\tilde V} W]^{2,0}=\alsq$ which in turn implies that
\bea
{\cal L}_V \omega_{(6)}= {\cal L}_{\tilde V} \omega_{(6)}=\alsq~.
\eea
Thus $M^6$ admits an almost Hermitian structure compatible with a connection $\hat{\tilde \nabla}^{(6)}$ with skew-symmetric torsion $W_{(6)}$.  It is well known
that in such case $W_{(6)}$ is determined in terms of the almost complex structure on $M^6$ and the metric, see eg \cite{howegp}.

To find whether $M^6$ inherits a $SU(3)$ structure as well, let investigate whether the (3,0) fundamental $SU(3)$ form $\chi$ of ${\cal S}$ descends on $M^6$.
It can always be arranged such that $i_V \chi=i_{\tilde V}\chi=0$.  So it remains to see whether $\chi$ is invariant under the action of $V$ and $\tilde V$.
For this a similar argument to that explained above for $\omega_{(6)}$ leads to the assertion that $\chi$ is invariant iff
the $\omega$-traces $i_{\tilde V} W\cdot \omega$ and $i_{\tilde V} W\cdot \omega $ of $i_{\tilde V} W$ and $i_{\tilde V} W$, respectively, vanish.  Furthermore,
an application of (\ref{bianx}) implies that both  $i_{\tilde V} W\cdot \omega$ and $i_{\tilde V} W\cdot \omega $ are constant but not necessarily zero.
Thus $M^6$ has generically a $U(3)$ structure instead of an $SU(3)$ one.

It remains to investigate the rest of the content of the conditions $\hat{\tilde \nabla} \tau_+^{(a)}=\alsq$.  First consider the (3,0) part of $W_{(6)}$
denoted by $W_{(6)}^{3,0}$. An application of (\ref{bianx}) using that $dW$ is a (2,2) form yields that
\bea
\hat{\tilde \nabla} W_{(6)}^{3,0}=\alsq~.
\eea
Thus $W_{(6)}^{3,0}$ is another globally defined $\hat{\tilde \nabla}$-parallel (3,0)-form on ${\cal S}$ and so it can either be set to zero or be identified with $\chi$.  In the
former case, the complex structure on $M^6$ is integrable and so $M^6$ is a KT manifold \cite{hkt}.

Writing $h=\lambda_1 V+ \lambda_2 \tilde V+h^\perp$, where $h^\perp$ is orthogonal to both $V$ and $\tilde V$ and $\lambda_1$ and $\lambda_2$ are constants, we find using (\ref{bianx}) that
\bea
\hat{\tilde\nabla} \big(2 d\Phi+ h^\perp-\theta_{\omega_{(6)}}\big)=\alsq~.
\label{covtheta}
\eea
Now if $2 d\Phi+ h^\perp-\theta_{\omega_{(6)}}$ in non-vanishing and since it is orthogonal to $V$ and $\tilde V$ will give rise to more than two $\eta_+$ $\tnp$-parallel
 spinors
on ${\cal S}$.  Since we have assumed that there are just two, we set
\bea
2 d\Phi+ h^\perp-\theta_{\omega_{(6)}}=\alsq~.
\eea
This concludes the investigation of geometry.

\subsection{Nearly supersymmetric horizons with $SU(2)$ holonomy}

\subsubsection{Assumptions and definitions}

It is known that if one requires the existence of an additional $\hat{\tilde\nabla}$-parallel spinor $\eta_+$ to those of the $SU(3)$ backgrounds on ${\cal S}$, then the isotropy algebra
of the all the five spinors reduces to $\mathfrak{su}(2)$.  As a result, ${\cal S}$ admits 8 $\hat{\tilde\nabla}$-parallel spinors and the holonomy group reduces to a subgroup $SU(2)$.
To describe the geometry of backgrounds with exactly 8 such spinors, we consider four linearly independent spinors  $\eta^{(a)}_+$,  and impose the condition
\bea
\label{grav_SU(2)}
\tnp \eta^{(a)}_+ = \alsq, \qquad a=0,1,2,3 \ ,
\eea
for which $\big({\cal{A}} \eta^{(a)}_+\big)^{[0]} \neq 0$, ($a=0, 1, 2, 3$).
It follows that the
 horizon section ${\cal{S}}^{[0]}$ admits a $SU(2)$ structure at zeroth order in $\alpha'$. We continue by setting   $\tau^{(a)}_+ = \cA \eta^{(a)}_+$.  These are non-vanishing and   satisfy
\bea
\label{gravtausu2}
\tnp \tau^{(a)}_+ = \alsq, \qquad a=0,1,2,3 \ .
\eea
Furthermore, we also define 1-form and 2-form spinor bilinears $V^{(a)}$ and $\omega_r$, respectively,
by
\bea
V_i\equiv V^{(0)}_i = \langle \eta^{(0)}_+ , \G_i \tau^{(0)}_+ \rangle, \qquad
(\omega_r)_{ij} = \langle \eta^{(0)}_+ , \G_{ij} \eta^{(r)}_+ \rangle~,~~~r=1,2,3~,
\eea
and also let
\bea
\tilde{V}_r = i_{V} \omega_r~.
\eea
In fact $\omega_r$ together with the metric and $W$ define an almost  HKT structure \cite{hkt} on ${\cal S}$  as
\bea
\label{par_condsu2}
\tnp V = \alsq , \qquad \tnp \omega_r = \alsq , \qquad \tnp \tilde{V}_r = \alsq \ ,
\eea
and the almost complex structures associated to $\omega_r$ satisfy the algebra of unit quaternions.
These  follow from  (\ref{naeta}) and the $\mathfrak{su}(2)$ isotropy of the parallel spinors.

\subsubsection{Symmetries of the horizon section}

It is clear from (\ref{par_condsu2}) that $V^{(a)}$, $V^{(r)}=V_r$,  generate isometries on ${\cal S}$ and that
\bea
i_aW=dV^{(a)}+\alsq~,
\eea
where $i_a$ denotes inner-derivation with respect to $V^{(a)}$.  Without loss of generality we choose $g(V^{(a)}, V^{(b)})=\ell^2 \delta^{ab}+\alsq$
for $\ell$ constant.
An investigation similar to the one explained in section \ref{symsu3} reveals that
\bea
&&{\cal L}_a \Phi=\alsq~,~~~{\cal L}_a W=\alsq~,~~~{\cal L}_a h=\al~,~~~i_adh=\al~,~~~
\cr
&&i_aF=\al~.
\eea
Next let us consider the commutator $[V^{(a)}, V^{(b)}]=i_a i_b W$. An application of (\ref{bianx}) together with the conditions above reveal that
\bea
\hat{\tilde\nabla}[V^{(a)}, V^{(b)}]=\alsq~.
\eea
Thus the commutator is either linear dependent on $V^{(a)}$ or it will lead to further reduction  of the holonomy of $\hat{\tilde\nabla}$ to $\{1\}$.
In the latter case, the horizon section ${\cal S}$ will admit more than four $\eta_+$ $\tnp$-parallel   spinors violating our assumptions. Thus, we conclude that
\bea
[V^{(a)}, V^{(b)}]=f^{ab}{}_c V^{(c)}+\alsq~,
\eea
for some constants $f$ with $\ell^2 f^{ab}{}_c= i_ai_bi_c W+\alsq$. As $f$ is skew-symmetric, the Lie algebra spanned by $V^{(a)}$ is a metric (compact) Lie algebra.
As it has dimension 4,  it is either isomorphic to $\oplus^4\mathfrak{u}(1)$ or to $\mathfrak{u}(1)\oplus \mathfrak{su}(2)$.

Therefore the horizon section ${\cal S}$ can be viewed locally as a fibration with fibre either $\times^4U(1)$ or $U(1)\times SU(2)$ over the space of orbits $M^4$ of  $V^{(a)}$.
We shall determine the geometry of ${\cal S}$ by specifying the geometry of $M^4$.

\subsubsection{ Geometry}

To simplify the analysis, we choose up to an $\mathfrak{so}(4)$ rotation $V$ to be along a $\mathfrak{u}(1)$ direction in either
$\oplus^4\mathfrak{u}(1)$ or $\mathfrak{u}(1)\oplus \mathfrak{su}(2)$.  This in particular implies that $i_0 i_r W=\alsq$. Then the
metric and torsion of ${\cal S}$ can be written as
\bea
d\tilde s^2=\ell^{-2} \delta_{ab} V^{(a)}\otimes V^{(b)}+d\tilde s^2_{(4)}+\alsq~,~~~ W=\ell^{-2} V\wedge dV+CS(V_r)+ W_{(4)}+\alsq
\nonumber \\
\eea
where $V^{(a)}$ is viewed as a principal bundle connection and $CS(V_r)$ is the Chern-Simons form which for the $\oplus^4\mathfrak{u}(1)$ case is
\bea
CS(V_r)=\ell^{-2}\sum_r V_r\wedge dV_r~.
\eea
The data $(ds^2_{(4)}, W_{(4)})$ define a geometry on $M^4$ with skew-symmetric torsion.

First, let us investigate the reduction of the almost HKT structure of ${\cal S}$ on $M^4$. For this observe that
\bea
\omega_r= \ell^{-2} V\wedge V_r+{\ell^{-2}\over2} \epsilon_r{}^{st}V_s\wedge V_t+\omega_r^{(4)}+\alsq~,
\eea
where $i_a \omega_r^{(4)}=\alsq$. Next consider ${\cal L}_a \omega^{(4)}_r$.  As both $V^{(a)}$ and $\omega^{(4)}_r$ are $\hat{\tilde \nabla}$-parallel, ${\cal L}_a \omega^{(4)}_r$
is specified by the properties of the rotation $i_a W$.  In particular if $i_a W$  is invariant under $\omega^{(4)}_r$,  the Lie derivative vanishes.

Next let us investigate the two cases $\oplus^4\mathfrak{u}(1)$ and $\mathfrak{u}(1)\oplus \mathfrak{su}(2)$
separately. In the abelian case, as $i_ai_bW=\alsq$, $i_a W$ is a 2-form on $M^4$. Furthermore ${\cal L}_a \omega^{(4)}_r$ vanishes iff the self-dual part,  $i_a W^{\rm sd}$, of $i_aW$ is zero. However in general
this may not be the case. An application of (\ref{bianx}) implies that
\bea
\hat{\tilde \nabla} i_a W^{\rm sd}=\alsq~,
\eea
and so  there exist some constants $u$ such that
\bea
i_a W^{\rm sd}= u_a{}^r \omega_r^{(4)}+\alsq~,
\eea
otherwise the holonomy of $\hat{\tilde \nabla}$ will be reduced further and it will admit more than four $\eta_+$ parallel spinors.
Then
\bea
{\cal L}_a \omega_r^{(4)}=2u_a{}^s\epsilon_{sr}{}^t \omega^{(4)}_t+\alsq~.
\eea
The identity $[{\cal L}_a, {\cal L}_b]={\cal L}_{[V^{(a)}, V^{(b)}]}$ gives
\bea
 (u_a^r u_b^s-u_b^r u_a^s)=\alsq~.
%{1\over2}f^{ab}{}_c u_c^t \epsilon_{tr}{}^s
\eea
The covariant constancy condition on $M^4$ now reads
\bea
\hat{\tilde \nabla}^{(4)} \omega^{(4)}_r=2 \ell^{-2} V^{(a)} u_a^s \epsilon_{sr}{}^t \omega^{(4)}_t+\alsq~,
\eea
where now $V^{(a)}$  should be thought as the pull back of the principal bundle connection $V^{(a)}$ with a local section.
It is clear that the relevant connection that determines the geometry of $M^4$ is $ Z^s=V^{(a)} u_a^s$.

If $u_a^r=0$, $M^4$ is a HKT manifold.  It is easy to see this as $\omega_r$ are covariantly constant with respect to a connection with skew-symmetric torsion
and  all three almost complex structures are integrable. The latter follows because  of dimensional reasons.  Otherwise one of the 3-vectors $u_a$ must be non-zero. Without
loss of generality take $u_0\not=0$.  In such a case the above equation can be solved as $(u_a^r)=(u_0^r, u_0^r v_s)$, where $v_s=|u_0|^{-2} \sum_r u^r_s u_0^r $.

Using these data, the covariant constancy condition of $\omega^{(4)}_r$ on $M^4$ can be written as
\bea
\hat{\tilde \nabla}^{(4)} \omega^{(4)}_r=2\ell^{-2} (V^0+V^p v_p) u_0^s \epsilon_{sr}{}^t \omega^{(4)}_t+\alsq~.
\eea
It is clear from this that $M^4$ is a KT manifold with respect to the Hermitian form $|u_0|^{-1} u_0^r \omega_r$. In fact $M^4$ is an (almost)\footnote{In the definition
of QKT structure in \cite{qkt} an additional integrability condition was considered.} QKT manifold \cite{qkt}
for which the holonomy of the $Sp(1)$ connection has been reduced to $U(1)$.

Next let us turn to examine the non-abelian $\mathfrak{u}(1)\oplus \mathfrak{su}(2)$ case. It is easy to see that
\bea
 (u_a^r u_b^s-u_b^r u_a^s)={1\over2}f^{ab}{}_c u_c^t \epsilon_{tr}{}^s+\alsq~.
\eea
If the 3-vector $u_0\not=0$, then all the rest of the components of $u$ vanish. In such a case, $M^4$ is an KT manifold. This class of solutions
includes the WZW type of solution $AdS_3\times S^3\times M^4$ where $M^4=S^1\times S^3$ with the bi-invariant metric and constant dilaton.
Such a horizon is not supersymmetric but it is nearly supersymmetric.

It remains to consider the case $u_0=0$. One can then show that $\det u\not=0$ and so $(u^r_s)$ is invertible. Thus $Z^s= V^{(a)} u_a^s$ takes values
in the $\mathfrak{sp}(1)$ Lie algebra.  $M^4$ is a QKT manifold, see also \cite{compgp}.

To conclude we remark that in all HKT and KT cases, there is an analogue of the condition (\ref{covtheta}) for every Hermitian form $\omega_r$
that determines these structures.  If the associated $2d\Phi+h^\perp-\theta_r$ forms do not vanish, then the holonomy of the connection with
torsion reduces to $\{1\}$ and the number of parallel spinors enhance to 16.  The solutions are the group manifolds. The solution $AdS_3\times S^3\times S^3\times S^1$ mentioned above
belongs to the class where the holonomy of the connection with torsion is $\{1\}$.

 There is an analogue
of this in the QKT case but in such a case the condition from the perspective of $M^4$ twists with $\mathfrak{sp}(1)$. If $2d\Phi+h^\perp-\theta_r$  do not vanish,
again the holonomy  of the connection with torsion on ${\cal S}$  reduces to $\{1\}$.  However now some of the data like the Hermitian forms
are not (bi-)invariant under the action of the group.  It would be of interest to explore his further to see whether there are actual solutions.

We conclude the examination of the geometry of nearly supersymmetric backgrounds in the $G_2$, $SU(3)$ and $SU(2)$ cases by pointing out that they exhibit an $\mathfrak{sl}(2,\bR)$ up to order $\al$ but not up to  order $\alsq$.
For the latter, $h$ must be a symmetry of the theory up to the same order and so it can be identified with $V$. The description of the
geometry  of this special class of nearly supersymmetric backgrounds is very similar to the one we have given above.  The only difference
is that now we can identify $h$ with $V$.

\newsection{Conclusions}

We have investigated the supersymmetric
 near-horizon geometry of heterotic black holes up to and including two loops in sigma model perturbation theory.
 Using a combination of
local and global techniques, together with the bosonic field equations and Bianchi identities, we have proven that the conditions
obtained from the KSEs are equivalent to a pair of gravitino equations  ({\ref{gravsimp}}) and a pair of algebraic conditions, related to the dilatino  KSE,
({\ref{algsimpmax}}), which are required to hold at zeroth and first order
in $\alpha'$. In particular, we have shown that the  KSE related to the gaugino
 is implied by the other KSEs and  field equations.

In all cases, we have also shown that
there are no regular $AdS_2$ solutions with compact without boundary internal space by demonstrating that $\Delta=\alsq$.
This is not in contradiction  with the fact that one can locally
write $AdS_3$ as a warped product over $AdS_2$ \cite{strominger}, see also appendix E. This is because our assumptions
on the internal space of $AdS_2$ are violated in such a case.

Furthermore, we have demonstrated that horizons that admit a non-vanishing $\eta_-$ Killing spinor up to order $\alsq$, which does not vanish at zeroth order in $\alpha'$, exhibit
supersymmetry enhancement  via
the same mechanism as described in \cite{hethor}, and so preserve 2, 4, 6 and 8 supersymmetries.  We have described the geometry
of such horizons in all cases and this is similar to that presented in \cite{hethor} for the horizons with $dH=0$.

We have also considered in some detail
the global properties of our solutions. The analysis of the global properties of $h^2$ proceeds in much the same way
as in the  heterotic theory with $dH=0$. However in the presence of anomaly, the consequences of the global restrictions on the geometry of the horizons
 are somewhat weaker. For example, it is only possible to prove that $h$ is an isometry
of the horizon section to zeroth order in $\alpha'$. So one cannot establish a direct algebraic relation
between $\eta_+$ and $\eta_-$ spinors to order $\alsq$, and therefore it is not
possible to directly show that there is supersymmetry enhancement via
this mechanism, as was done  in \cite{hethor} for the theory with $dH=0$.

We have also constructed generalized Lichnerowicz
type theorems, which relate spinors which are parallel with respect to a certain type of near-horizon supercovariant derivative,
to zero modes of near-horizon Dirac operators.
We have shown that if $\eta$ is a zero mode of the near-horizon
Dirac operator to both zero and first order in $\alpha'$,
then the Lichnerowicz theorems imply that $\eta$ only satisfies
the KSE ({\ref{gravsimp}}) and
({\ref{algsimpmax}}) to zero order in $\alpha'$. Hence, the
types of arguments used to show supersymmetry enhancement via
Lichnerowicz type theorems
in \cite{lichner11, lichneriib, lichneriia1, lichneriia2} also do
not work to the required order in $\alpha'$ for the heterotic theory.

Finally, we have examined a class of nearly supersymmetric horizons for which the gravitino KSE is allowed
to admit solutions on the spatial horizon section but not the rest of the KSEs.  Such solutions
in general do not admit any spacetime Killing spinors including solutions of the gravitino KSE.
Under some conditions on the fluxes, we investigate the geometry of the spatial horizon sections
using a combination of local and global techniques as well as the field equations.
We find that those with a $G_2$, $SU(3)$ and $SU(2)$ structure admit 1, 2 and 4 parallel vectors
on the spatial horizon sections with respect to the connection with torsion. The geometry on the orbit
spaces of these isometries is fully specified.

The spacetime of both supersymmetric, and nearly supersymmetry
horizons considered here admits a $SL(2,\bR)$ symmetry at zeroth order in $\alpha'$.
In the supersymmetric case for which there is a $\eta_-$ Killing spinor to order $\alsq$ such that  $\eta_-$ does not vanish at zeroth order, $\eta_-^{[0]}\not=0$,
this symmetry persists at first order in $\alpha'$.
The nearly supersymmetric horizons also admit an $SL(2,\bR)$ symmetry provided that $h$ is parallel with respect to the connection with torsion up to $\alsq$.

It is not apparent whether the properties of the heterotic horizons described here are going to persist to higher than two loops in sigma model perturbation theory.
It is likely though that the presence of an $\mathfrak{sl}(2,\bR)$ symmetry will persist after perhaps a suitable choice of a scheme in perturbation theory. There is no apparent reason
to hypothesize that such a symmetry can be anomalous at higher loops.
What happens to global properties of the horizons, for example
the Lichnerowicz type theorems, is less clear.
We have already seen that these theorems do not hold to the expected order in $\alpha'$ even at two loops. This can be taken as an indication
that additional higher order corrections may further weaken the consequences of such theorems.

\vskip 0.5cm
\noindent{\bf Acknowledgements} \vskip 0.1cm
\noindent  AF is partially supported by the EPSRC grant FP/M506655. JG is supported by the STFC grant, ST/1004874/1.  GP is partially supported by the  STFC rolling grant ST/J002798/1.

\vskip 0.5cm

\vskip 0.5cm
%\newpage
\noindent{\bf Data Management} \vskip 0.1cm

\noindent No additional research data beyond the data presented and cited in this work are
needed to validate the research  findings in this work.

\vskip 0.5cm

\newpage

\setcounter{section}{0}
\setcounter{subsection}{0}

\appendix{Useful formulae}

\subsection{Spin Connection and Curvature}

In our conventions, the curvature of a connection $\Gamma$ is given by
\bea
R_{AB,}{}^C{}_D = \partial_A \Gamma^C_{BD}-\partial_B \Gamma^C_{AD} + \Gamma^C_{AN} \Gamma^N_{BD}-\Gamma^C_{BN}
\Gamma^N_{AD}~,
\eea
We define connections $\check \nabla$ and $\hat \nabla$
as follows
\bea
\label{torsionconv}
{\check \nabla}_M \xi^N = \nabla_M \xi^N -{1 \over 2} H^N{}_{ML} \xi^L,
\qquad
{\hat \nabla}_M \xi^N = \nabla_M \xi^N +{1 \over 2} H^N{}_{ML} \xi^L
\eea
for vector field $\xi$, where $\nabla$ is the Levi-Civita connection.
In particular, the $\check R$ curvature tensor can be written as
\bea
\check R_{AB,CD} = { R}_{ABCD}
-{1 \over 2} \nabla_A H_{CBD}+{1 \over 2} \nabla_B H_{CAD}
+{1 \over 4} H_{CAN} H^N{}_{BD} -{1 \over 4} H_{CBN}H^N{}_{AD}~,
\eea
where $ { R}$ is the Riemann curvature tensor. Also, note that
\bea
\label{curvcross}
\check R_{AB,CD}-\hat R_{CD,AB}= {1 \over 2}(dH)_{ABCD} \ .
\eea

We also define connections ${\check {\tilde \nabla}}$ and ${\hat {\tilde \nabla}}$ on the horizon section ${\cal S}$ via
\bea
\label{torsionconv2}
{\check {\tilde \nabla}}_i Y^j = {\tilde{\nabla}}_i Y^j -{1 \over 2} W^j{}_{ik} Y^k, \qquad {\hat {\tilde \nabla}}_i Y^j = {\tilde{\nabla}}_i Y^j +{1 \over 2} W^j{}_{ik} Y^k~,
\eea
for vector fields $Y$ on ${\cal{S}}$, and where ${\tilde \nabla}$ is the Levi-Civita connection of ${\cal S}$, and we denote the curvatures
of the connections ${\tilde{\nabla}}$, ${\check {\tilde{\nabla}}}$ and
${\hat {\tilde{\nabla}}}$ by ${\tilde{R}}$, ${\check {\tilde{R}}}$ and
${\hat {\tilde{R}}}$ respectively.

The non-vanishing components of the spin connection in
the frame basis ({\ref{nhbasis}}) of the near horizon metric (\ref{nearhormetr})  are
\begin{eqnarray}
&&\Omega_{-,+i} = -{1 \over 2} h_i~,~~~
\Omega_{+,+-} = -r \Delta, \quad \Omega_{+,+i} ={1 \over 2} r^2(  \Delta h_i - \partial_i \Delta),
\cr
&&\Omega_{+,-i} = -{1 \over 2} h_i, \quad \Omega_{+,ij} = -{1 \over 2} r dh_{ij}~,~~~
\Omega_{i,+-} = {1 \over 2} h_i, \quad \Omega_{i,+j} = -{1 \over 2} r dh_{ij},
\cr
&&\Omega_{i,jk}= \tilde\Omega_{i,jk}~,
\end{eqnarray}
where $\tilde\Omega$ denotes the spin-connection of the spatial horizon section  ${{\cal{S}}}$ in the   ${\bf{e}}^i$ basis.
If $f$ is any function of spacetime, then frame derivatives are expressed in terms of co-ordinate derivatives  as
\begin{eqnarray}
\partial_+ f &=& \partial_u f +{1 \over 2} r^2 \Delta \partial_r f~,~~
\partial_- f = \partial_r f~,~~
\partial_i f = {\tilde{\partial}}_i f -r \partial_r f h_i \ .
\end{eqnarray}
The non-vanishing components of the Ricci tensor is the
 basis ({\ref{nhbasis}}) are
\bea
R_{+-} &=& {1 \over 2} \hn^i h_i - \Delta -{1 \over 2} h^2~,~~~
R_{ij} = {\tilde{R}}_{ij} + \hn_{(i} h_{j)} -{1 \over 2} h_i h_j
\nonumber \\
R_{++} &=& r^2 \big( {1 \over 2} \hn^2 \Delta -{3 \over 2} h^i \hn_i \Delta -{1 \over 2} \Delta \hn^i h_i + \Delta h^2
+{1 \over 4} (dh)_{ij} (dh)^{ij} \big)
\nonumber \\
R_{+i} &=& r \big( {1 \over 2} \hn^j (dh)_{ij} - (dh)_{ij} h^j - \hn_i \Delta + \Delta h_i \big) \ ,
\eea
where ${\tilde{R}}$ is the Ricci tensor of the horizon section ${\cal S}$ in the $\bbe^i$ frame.

We remark that the non-vanishing components of the Hessian of $\Phi$, are given by
\bea
\label{hessian1}
\nabla_+ \nabla_- \Phi &=& -{1 \over 2} h^i \hn_i \Phi~,
\nonumber \\
\nabla_+ \nabla_i \Phi &=& -{1 \over 2} r (dh)_i{}^j \hn_j \Phi~,
\nonumber \\
\nabla_i \nabla_j \Phi &=& \hn_i \hn_j \Phi~,
\eea
where in the above expression, we have set $\Delta=0$.

The non-vanishing components of the $\check R$ curvature tensor in the basis ({\ref{nhbasis}}) are
\bea
\label{rmin}
\check R_{-i,+j} &=& \hn_j h_i +{1 \over 2} h^\ell W_{\ell ij}
~,~~~
\check R_{ij,+-} = dh_{ij}~,
\nonumber \\
\check R_{ij,+k} &=& r \bigg( \hn_k dh_{ij} -h_k dh_{ij}
+{1 \over 2} (dh)_i{}^m W_{mjk} -{1 \over 2} (dh)_j{}^m W_{mik} \bigg)~,
\nonumber \\
\check R_{ij,k\ell} &=& {\tilde{R}}_{ijk\ell}- {1 \over 2}\hn_i W_{kj\ell}
+{1 \over 2} \hn_j W_{ki \ell} +{1 \over 4} W_{kim} W^m{}_{j \ell}
-{1 \over 4} W_{kjm} W^m{}_{i \ell}
\nonumber
\\
&=& \check {\tilde{R}}_{ij,k \ell}~,
\eea
where in the above expression, we have set $\Delta=0$, $N=h$ and $Y=dh$.
Note that the $\check R_{-i,+j}$ and $\check R_{ij,+k}$ terms give no
contribution to the Bianchi identity of $H$ or to the Einstein equations,
because $\check R_{MN,-i}=0$ for all $M,N$.

\subsection{Bosonic Field Equations}

The Bianchi identity associated with the 3-form is
\bea
\label{bian}
dH = - {\alpha' \over 4} \bigg( {\rm tr}(\check R \wedge \check R) - {\rm tr}(F \wedge F) \bigg) + \alsq
\eea
where ${\rm tr} (F \wedge F) = F^a{}_b \wedge F^b{}_a$ ($a, b$ are gauge indices on $F$).

The Einstein equation is
\bea
\label{ein}
&&R_{MN} -{1 \over 4} H_{M L_1 L_2} H_N{}^{L_1 L_2}
+2 \nabla_M \nabla_N \Phi
\cr &&~~~~~~~
+ {\alpha' \over 4} \bigg(\check R_{M L_1, L_2 L_3}
\check R_N{}^{L_1, L_2 L_3}-F_{M Lab}F_N{}^{Lab} \bigg)=\alsq~.
\nonumber \\
\eea
The gauge field equations are
\bea
\label{geq1}
\nabla^M \bigg(e^{-2 \Phi} H_{M N_1 N_2}\bigg)= \alsq~,
\eea
and
\bea
\label{geq2}
\nabla^M \bigg(e^{-2 \Phi}F_{MN} \bigg)+{1 \over 2} e^{-2 \Phi} H_{NL_1 L_2} F^{L_1 L_2}= \al~.
\eea
The dilaton field equation is
\bea
\label{deq}
\nabla_M \nabla^M \Phi &=& 2 \nabla_M \Phi \nabla^M \Phi -{1 \over 12} H_{N_1 N_2 N_3} H^{N_1 N_2 N_3}
\nonumber \\
&+& {\alpha' \over 16} \bigg(\check R_{N_1 N_2, N_3 N_4}
\check R^{\ N_1 N_2, N_3 N_4}-F_{N_1 N_2 ab}F^{N_1 N_2 ab} \bigg) + \alsq \ .
\eea
This completes the list of field equations. We have followed the conventions of \cite{tsimpis}.

\appendix{Further Simplification of the KSEs}
\setcounter{subsection}{0}

\label{ind}

Here we shall show that the independent  KSEs are given in (\ref{gravsimp}) and (\ref{algsimpmax}).
We first note that the  conditions on the bosonic fields  ({\ref{bossimp1}})
(obtained from the case when $\phi_+^{[0]} \equiv 0$)
actually imply those of  ({\ref{bossimp2}})
(corresponding to the $\phi_+^{[0]} \not \equiv 0$ case). Furthermore, the KSEs ({\ref{par3bb}}), ({\ref{auxalg1bbb}}),
({\ref{auxalg1cbb}}), ({\ref{par4bb}}), ({\ref{auxalg2bbb}})
and ({\ref{auxalg2cbb}}) are identical to
the KSE ({\ref{par3}}), ({\ref{auxalg1b}}), ({\ref{auxalg1c}}), ({\ref{par4bb}}), ({\ref{auxalg2b}})
and ({\ref{auxalg2c}}).
Hence, we shall concentrate on the simplification of the KSEs associated
with the case $\phi_+^{[0]} \not \equiv 0$, as the simplification of
the KSEs in the case $\phi_+^{[0]} \equiv 0$ follows in exactly the same way.

\subsection{Elimination of conditions ({\ref{auxalg1}}),
({\ref{auxalg1c}}), ({\ref{auxalg2}}), ({\ref{auxalg2c}})}

Let us assume ({\ref{par3}}), ({\ref{auxalg1b}}),
({\ref{par4}}) and ({\ref{auxalg2b}}). Then acting on the algebraic conditions
({\ref{auxalg1b}}) and ({\ref{auxalg2b}}) with the Dirac operator
$\Gamma^\ell \hn_\ell$, one obtains
\bea
\bigg(\hn_i \hn^i \Phi \mp h^i \hn_i \Phi -2 \hn^i \Phi \hn_i \Phi
-{1 \over 2} h_i h^i +{1 \over 12} W_{ijk} W^{ijk}
\nonumber \\
+{1 \over 4} (1\pm1) dh_{ij} \Gamma^{ij} +{1 \over 4} (-1\pm 1)h^k W_{kij}
\Gamma^{ij} -{1 \over 48} dW_{\ell_1 \ell_2 \ell_3 \ell_4}
\Gamma^{\ell_1 \ell_2 \ell_3 \ell_4} \bigg) \phi_\pm = \alsq~,
\eea
where we have made use of the field equations ({\ref{geq1a}}) and
({\ref{geq1c}}), together with the algebraic conditions
({\ref{auxalg1b}}) and ({\ref{auxalg2b}}).
Next, on substituting the dilaton equation and the Bianchi identity
into the above expression, one finds
\bea
\label{reduc1}
\bigg({1 \over 2}(1 \mp 1) \hn^i h_i
+{1 \over 4} (1 \pm 1) dh_{ij} \Gamma^{ij}
+{1 \over 4}(-1 \pm 1) (i_h W)_{ij} \Gamma^{ij}
\nonumber \\
+{\alpha' \over 16} dh_{ij} \Gamma^{ij} dh_{pq} \Gamma^{pq}
+{\alpha' \over 32} {\tilde{F}}_{ij}{}^{ab} \Gamma^{ij}
{\tilde{F}}_{pq ab}\Gamma^{pq}
-{\alpha' \over 32} \check{\tilde{R}}_{ij,}{}^{mn}
\Gamma^{ij} \check {\tilde{R}}_{pq,mn} \Gamma^{pq} \bigg) \phi_\pm
=\alsq~.
\eea
Further simplification can be obtained by noting that
the integrability conditions of the KSE ({\ref{par3}}) and ({\ref{par4}})
are
\bea
\hat{\tilde{R}}_{ij,pq} \Gamma^{pq} \phi_\pm = \alsq~,
\eea
and hence
\bea
\check{\tilde{R}}_{pq,ij} \Gamma^{pq} \phi_\pm = \al~,
\eea
from which it follows that the final term on the RHS of
({\ref{reduc1}}) is $\alsq$ and hence can be neglected.
So, ({\ref{reduc1}}) is equivalent to
\bea
\label{reduc1b}
\bigg({1 \over 2}(1 \mp 1) \hn^i h_i
+{1 \over 4} (1 \pm 1) dh_{ij} \Gamma^{ij}
+{1 \over 4}(-1 \pm 1) (i_h W)_{ij} \Gamma^{ij}
\nonumber \\
+{\alpha' \over 16} dh_{ij} \Gamma^{ij} dh_{pq} \Gamma^{pq}
+{\alpha' \over 32} {\tilde{F}}_{ij}{}^{ab} \Gamma^{ij}
{\tilde{F}}_{pqab}\Gamma^{pq} \bigg) \phi_\pm
=\alsq~.
\eea

We begin by considering the condition which ({\ref{reduc1b}}) imposes
on $\phi_+$:
\bea
\label{reduc1c}
\bigg({1 \over 2} dh_{ij} \Gamma^{ij}
+{\alpha' \over 16} dh_{ij} \Gamma^{ij} dh_{pq} \Gamma^{pq}
+{\alpha' \over 32} {\tilde{F}}_{ij}{}^{ab} \Gamma^{ij}
{\tilde{F}}_{pqab}\Gamma^{pq} \bigg) \phi_+
=\alsq
\eea
To zeroth order this gives
\bea
dh_{ij} \Gamma^{ij} \phi_+ = \al~,
\eea
which implies that the second term on the LHS of ({\ref{reduc1c}}) is
of $\alsq$, and hence can be neglected. Using this,
({\ref{reduc1c}}) gives that
\bea
\alpha' \langle {\tilde{F}}_{ij}{}^{ab} \Gamma^{ij} \phi_+,
{\tilde{F}}_{pqab} \Gamma^{pq} \phi_+ \rangle = \alsq~,
\eea
which implies that
\bea
{\tilde{F}}_{ij}{}^{ab} \Gamma^{ij} \phi_+ = \al \ .
\eea
Using this the third term on the LHS of ({\ref{reduc1c}})
is also of $\alsq$. So, the remaining content
of ({\ref{reduc1c}}) is
\bea
dh_{ij} \Gamma^{ij} \phi_+ = \alsq \ .
\eea
Hence, we have proven that the KSE ({\ref{par3}}) and ({\ref{auxalg1b}})
imply the algebraic KSE ({\ref{auxalg1}}) and  ({\ref{auxalg1c}}).

Next, we consider the condition which  ({\ref{reduc1b}}) imposes
on $\phi_-$, which is
\bea
\label{reduc1d}
\bigg(\hn^i h_i -{1 \over 2} (i_h W)_{ij} \Gamma^{ij}
+{\alpha' \over 16} dh_{ij} \Gamma^{ij} dh_{pq} \Gamma^{pq}
+{\alpha' \over 32} {\tilde{F}}_{ij}{}^{ab} \Gamma^{ij}
{\tilde{F}}_{pqab}\Gamma^{pq} \bigg) \phi_-
=\alsq~.
\eea
However, note also that the $u$-dependent part of
({\ref{par3}}), with ({\ref{par4}}), implies that
\bea
\label{udep1}
\bigg(\hn_i h_j - {1 \over 2} W_{ijk} h^k \bigg) \Gamma^j \phi_-= \alsq~.
\eea
On contracting this expression with $\Gamma^i$,
we find
\bea
\label{udep1b}
\bigg(\hn^i h_i +{1 \over 2} dh_{ij}\Gamma^{ij} -{1 \over 2}
(i_h W)_{ij} \Gamma^{ij} \bigg) \phi_- = \alsq~,
\eea
and on substituting this expression into ({\ref{reduc1d}}) we get
\bea
\label{reduc1e}
\bigg(-{1 \over 2} dh_{ij} \Gamma^{ij}
+{\alpha' \over 16} dh_{ij} \Gamma^{ij} dh_{pq} \Gamma^{pq}
+{\alpha' \over 32} {\tilde{F}}_{ij}{}^{ab} \Gamma^{ij}
{\tilde{F}}_{pqab}\Gamma^{pq} \bigg) \phi_-
=\alsq~.
\eea
Hence, we find from exactly the
same reasoning which was used to analyse the conditions on $
\phi_+$, that ({\ref{par4}}) and ({\ref{auxalg2b}}) imply
({\ref{auxalg2}}) and ({\ref{auxalg2c}}).

So, on making use of the  field equations, it follows that
the necessary and sufficient conditions for supersymmetry simplify
to the conditions ({\ref{par3}}) and ({\ref{auxalg1b}}) on
$\phi_+$, and to ({\ref{par4}}) and ({\ref{auxalg2b}}) on $\eta_-$.
We remark that the $u$-dependent parts of
the conditions ({\ref{par3}}) and ({\ref{auxalg1b}}) also impose
conditions on $\eta_-$. We shall examine the conditions on
$\eta_-$ further in the next section, and show how these may be
simplified.

\subsection{Elimination of $u$-dependent parts of ({\ref{par3}}) and
({\ref{auxalg1b}})}

We begin by considering the $u$-dependent parts of ({\ref{par3}}) and
({\ref{auxalg1b}}), assuming that ({\ref{par4}}) and ({\ref{auxalg2b}})
hold. The $u$-dependent part of the condition on $\phi_+$
obtained from ({\ref{par3}}) is
\bea
\label{udepa}
\bigg(\hn_i h_j - {1 \over 2} W_{ijk} h^k \bigg) \Gamma^j \eta_-= \alsq~,
\eea
and the $u$-dependent part of the algebraic condition
({\ref{auxalg1b}}) is given by
\bea
\label{udepb}
\bigg(\Gamma^i \hn_i \Phi +{1 \over 2} h_i \Gamma^i
-{1 \over 12} W_{ijk}\Gamma^{ijk} \bigg) h_\ell \Gamma^\ell \eta_- = \alsq~.
\eea
On adding $h_\ell \Gamma^\ell$ acting on ({\ref{auxalg2b}}) to the
above expression, we find that ({\ref{udepb}}) is equivalent to
the condition
\bea
\label{udepc}
\bigg(\hn^i h_i -{1 \over 2} h^i W_{ijk} \Gamma^{jk} \bigg) \eta_-~,
= \alsq
\eea
where we have also made use of the field equation ({\ref{geq1a}}).
On contracting ({\ref{udepa}}) with $\Gamma^i$, it then follows
that ({\ref{udepc}}) is equivalent
to
\bea
dh_{ij} \Gamma^{ij} \eta_- = \alsq \ .
\eea
However, as shown in the previous section, this condition
is implied by ({\ref{par4}}) and ({\ref{auxalg2b}})
on making use of the  field equations.

So, it remains to consider the condition ({\ref{udepa}}).
First, recall that the integrability conditions
of the gravitino equation of ({\ref{par4}}) is given
by
\bea
\hat{{\tilde{R}}}_{ij,k\ell} \Gamma^{k \ell} \eta_- = \alsq~.
\eea
On contracting with $\Gamma^j$, one then obtains
\bea
\label{minint1}
\bigg( \big(-2 {\tilde{R}}_{ij} +{1 \over 2} W_{imn}W_j{}^{mn}
-2 \hn^k \Phi W_{kij} + dh_{ij} - h^k W_{kij} \big) \Gamma^j
\nonumber \\
+ \big(-{1 \over 6} (dW)_{ijk\ell}-{1 \over 3} \hn_i W_{jk\ell}
+{1 \over 2} W_{ij}{}^m W_{k \ell m} \big) \Gamma^{jk\ell} \bigg)
\eta_- = \alsq~,
\eea
where we have used the gauge equation ({\ref{geq1c}}).
Also, on taking the covariant derivative of the algebraic condition
({\ref{auxalg2b}}), and using ({\ref{par4}}), one also finds the
following mixed integrability condition
\bea
\label{minint2}
\bigg( \big(\hn_i \hn_j \Phi -{1 \over 2} \hn_i h_j +{1 \over 2} W_{ikj}
\hn^k \Phi -{1 \over 4} W_{ikj} h^k \big) \Gamma^j
\nonumber \\
+\Gamma^{jk\ell} \big(-{1 \over 12} \hn_i W_{jk\ell}
+{1 \over 8} W_{jkm}W_{i\ell}{}^m \big) \bigg) \eta_-=\alsq~.
\eea
 On eliminating the $\hn_i W_{jk\ell} \Gamma^{jk\ell}$ terms
between ({\ref{minint1}}) and ({\ref{minint2}}), one obtains
the condition
\bea
\bigg(\big(-2 {\tilde{R}}_{ij} +{1 \over 2} W_{imn}W_j{}^{mn}
+dh_{ij} -2 h^k W_{kij} -4 \hn_i \hn_j \Phi
+2 \hn_i h_j \big) \Gamma^j
\nonumber \\
-{1 \over 6} dW_{ijk\ell} \Gamma^{jk\ell}
\bigg) \eta_-=\alsq~.
\eea
Next, we substitute the Einstein equation ({\ref{einsp}}) in order
to eliminate the Ricci tensor, and also use the Bianchi identity for $dW$.
One then obtains, after some rearrangement of terms, the following
condition
\bea
\label{minint3}
\bigg( \big(4 \hn_i h_j -2 h^k W_{kij} \big) \Gamma^j
+ \alpha' \big( {1 \over 2} dh_{ij} \Gamma^j dh_{k \ell} \Gamma^{k \ell}
+{1 \over 4} {\tilde{F}}_{ijab} \Gamma^j {\tilde{F}}_{k \ell}{}^{ab}
\Gamma^{k \ell}
\nonumber \\
 -{1 \over 4} \check{\tilde{R}}_{ij,}{}^{mn}
\Gamma^j \check{\tilde{R}}_{k \ell,mn} \Gamma^{k\ell} \big) \bigg)
\eta_- = \alsq~.
\nonumber \\
\eea
The $\alpha'$ terms in the above expression can be neglected, as they
all give rise to terms which are in fact $\alsq$.
This is because of the conditions ({\ref{auxalg2}}) and ({\ref{auxalg2c}}),
which we have already shown follow from ({\ref{par4}}) and ({\ref{auxalg2b}}), together with the bosonic conditions,
as well as the fact that
\bea
\check{\tilde{R}}_{k\ell, mn} \Gamma^{k \ell}\eta_-=\al~,
\eea
which follows from the integrability condition of ({\ref{par4}}).
It follows that ({\ref{minint3}}) implies ({\ref{udepa}}).

\appendix{A consistency condition}

Suppose that we consider
the Bianchi identity associated with the 3-form as
\bea
\label{bianx}
dH = - {\alpha' \over 4} \bigg( {\rm tr}({\cal R}  \wedge {\cal R}) - {\rm tr}(F \wedge F) \bigg) + \alsq~,
\eea
where $ {\cal R}$ is a spacetime curvature which will be specified later.
Also observe that the 2-form gauge potential and the Einstein equation can be written together as
\bea
\label{einx}
{\cal{E}}_{MN} \equiv \hat R_{MN}
+2\hat\nabla_M \nabla_N \Phi
+ {\alpha' \over 4} \bigg({\cal R}_{M L_1, L_2 L_3}
{\cal R}_N{}^{L_1, L_2 L_3}-F_{M Lab}F_N{}^{Lab} \bigg)=\alsq~.
\eea
Then one can establish by direct computation that
\bea
\hat R_{M[N,PQ]}=-{1\over3} \hat\nabla_M H_{NPQ}-{1\over6} dH_{MNPQ}~.
\label{bianx}
\eea
Using this condition, one can derive the relation
\bea
&&\hat R_{MN,PQ}\Gamma^N \Gamma^{PQ}\epsilon=-2 {\cal{E}}_{MQ} \Gamma^Q \epsilon -{1\over3} \hat\nabla_M\big(H_{LPQ} \Gamma^{LPQ}-12 \partial_L\Phi \Gamma^L\big)\epsilon
\cr
&&~~~~~~-{\alpha'\over4} [{\cal R}_{MN,EF} {\cal R}_{PQ,}{}^{EF}- F_{MN ab} F_{PQ}{}^{ab}] \Gamma^N \Gamma^{PQ}\epsilon + \alsq~.
\label{xxxid}
\eea
If $\epsilon$ satisfies that gravitino KSE, the left hand side of this relation vanishes.  Furthermore the right-hand-side vanishes
as well, provided that the Einstein and gauge field equations hold and the dilatino and gaugino KSEs are satisfied, and in addition
\bea
 {\cal R}_{PQ,}{}^{EF}\Gamma^{PQ}\epsilon=\al~.
\label{instcon}
\eea
Of course in heterotic string perturbation theory
\bea
\check R_{PQ,}{}^{EF}\Gamma^{PQ}\epsilon={\cal O}(\alpha')~,
 \eea
 as a consequence of the gravitino KSE and the closure of $H$ at that order. Thus one can set $ {\cal R}=\check R$ and the identity
 (\ref{xxxid}) will hold up to order $\alpha'^2$. 
 
 Clearly (\ref{xxxid}) arises from the integrability of the gravitino KSE and requires  
 a compatibility between the  KSEs and field equations of the theory. Similar identities can be derived from the remaining integrability conditions of KSEs  of the theory.  In particular there is an identity derived from the integrability condition  of dilatino and gravitino KSEs which contains   the dilaton field equation, including the
 2-loop correction, the $H$ field equation and $dH$. The compatibility between the field equations and this integrability condition is again resolved by taking 
 $F$ and ${\cal R}$ to satisfy the gaugino KSE as above.

 One consequence of the identity (\ref{xxxid}) is that if the gravitino KSE and gaugino KSEs are satisfied as well as (\ref{instcon}) but the dilatino is not, then
 the gravitino KSE admits an additional parallel spinor of the opposite chirality.  Such kinds of identities have been established before
 for special cases in \cite{howegp}. Here we have shown that such a result is generic in the context of heterotic theory.

\appendix{Lichnerowicz Theorem Computation}
\label{calcId}

In this appendix, we present the details for the calculation of
the functional ${\cal{I}}$ defined in ({\ref{I functional}}),
and show how the constants $q$ and $c$ are fixed by requiring
that certain types of terms which arise in the calculation should
vanish. We begin by considering the calculation at zeroth order
in $\alpha'$, and then include the corrections at first order in
$\alpha'$. We remark that we shall retain terms of the type
$h^i \hn_i \Phi$ throughout. This is because although
these terms vanish at zeroth order in $\alpha'$ as a consequence
of the analysis in Section 8, it does not follow from this analysis
that ${\cal{L}}_h \Phi = \alsq$. However, as we shall see, it turns out that the
coefficient multiplying the terms $h^i \hn_i \Phi$, which depends
on the  constants $q$ and $c$, vanishes when one requires that
several other terms in ${\cal{I}}$ vanish as well. So these
terms do not give any contribution to ${\cal{I}}$ at either zeroth or
first order in $\alpha'$.

\setcounter{subsection}{0}

\subsection{Computations at zeroth order in $\a'$}
\label{I zeroth order}
Throughout the following analysis, we assume Einstein equations, dilaton field equation and Bianchi identity at zeroth order in $\a'$.
To proceed, we expand out the definition of ${\nabla}^{(\kappa)}_i$ and $\cD$ in $\cI$, obtaining the following expression
\begin{align}
\label{expansion}
\notag
\cI &= \int_{\cS} e^{c\Phi} 2(\k - q) \langle \G^i \cA \eta_{\pm} , \tnp_i \eta_{\pm} \rangle + e^{c\Phi} (8\k^2 - q^2) \langle \eta_{\pm} , \cA^{\dagger} \cA \eta_{\pm} \rangle  \\
& \qquad\quad - e^{c\Phi} \langle \tnp_i \eta_{\pm} , \G^{ij} \tnp_j \eta_{\pm} \rangle \ .
\end{align}

 Now, after writing $\tnp$ in terms of the Levi-Civita connection $\tilde{\nabla}$ and after integrating by parts, the expression (\ref{expansion}) decomposes into
\be
\cI = \cI_1 +\cI_2 + \cI_3 \ ,
\ee
where
\begin{align}
\cI_1 = \int_{\cS}  &e^{c\Phi} 2(\k- q) \langle \eta_{\pm} , \cA^{\dagger} \cD \eta_{\pm} \rangle + e^{c\Phi} (8\k^2 - 2 \k q + q^2) \langle \eta_{\pm} , \cA^{\dagger}\cA \eta_{\pm} \rangle   \\
&- \frac{1}{64} e^{c\Phi}\langle \eta_{\pm} , \G^{\ell_1\ell_2} \G^{ij} \G^{\ell_3\ell_4} W_{i\ell_1\ell_2}W_{j\ell_3\ell_4} \eta_{\pm} \rangle  \ ,
\end{align}
and
\begin{align}
\notag
\cI_2 = \int_{\cS} &c e^{c\Phi} \langle \eta_{\pm} , \G^{ij} \tilde{\nabla}_j \eta_{\pm} \rangle  + \frac{1}{8} e^{c\Phi} \langle \tilde{\nabla}_i \eta_{\pm}, \G^{ij}\G^{\ell_1\ell_2} W_{j\ell_1\ell_2} \eta_{\pm} \rangle  \\
&- \frac{1}{8} e^{c\Phi} \langle \eta_{\pm}, \G^{\ell_1\ell_2} \G^{ij} W_{j\ell_1\ell_2} \tilde{\nabla}\eta_{\pm} \rangle  \ ,
\end{align}
and
\be
\cI_3 = \int_{\cS} - e^{c\Phi} \langle \tilde{\nabla}_i \eta_{\pm} , \G^{ij} \tilde{\nabla}_j \eta_{\pm} \rangle \ .
\ee
In particular, we note the identity
\be
\G^{\ell_1\ell_2} \G^{ij} \G^{\ell_3\ell_4} W_{i\ell_1\ell_2}W_{j\ell_3\ell_4} = 8 W^i{}_{\ell_1\ell_2}W_{i\ell_3\ell_4} \G^{\ell_1\ell_2\ell_3\ell_4} - 4 W_{ijk}W^{ijk} \ ,
\ee
which simplifies $\cI_1$. After integrating by parts the second term in $\cI_2$, we have
\begin{align}
\label{I_2 parts}
\notag
\cI_2 = \int_{\cS} &c e^{c\Phi} \langle \eta_{\pm} , \G^{ij} \tilde{\nabla}_j \eta_{\pm} \rangle  - \frac{1}{8} e^{c\Phi} \langle \eta_{\pm} , \left( \G^{ij}\G_{mn} - \G_{mn}\G^{ij} \right) W_j{}^{mn} \tn_i \eta_{\pm} \rangle \\
& - \frac{c}{8} e^{c\Phi} \langle \eta_{\pm} , \G^{i\ell_1\ell_2\ell_3} \tn_i \Phi W_{\ell_1\ell_2\ell_3} \eta_{\pm} \rangle    - \frac{1}{8}e^{c\Phi} \langle \eta_{\pm} , \G^{\ell_1\ell_2\ell_3\ell_4} \tn_{\ell_1}W_{\ell_2\ell_3\ell_4} \eta_{\pm} \rangle  \ ,
\end{align}
where the last term is order $\alpha'$, so we shall neglect it. Now we shall focus on the second term of (\ref{I_2 parts}). First note that
\be
\left( \G^{ij}\G_{mn} - \G_{mn}\G^{ij} \right) W_j{}^{mn} = - 4 \G^{mn} W^i{}_{mn} =  \frac{4}{3} W_{\ell_1\ell_2\ell_3} \left( \G^{\ell_1\ell_2\ell_3}\G^i + \G^{i\ell_1\ell_2\ell_3} \right) \ .
\ee
Then, after an integration by parts and after writing $\tn$ in terms of $\cD$, we have
\begin{align}
\label{GGexp}
\notag
\int_{\cS}- \frac{1}{8} e^{c\Phi} \langle \eta_{\pm} , &\left( \G^{ij}\G_{mn} -  \G_{mn}\G^{ij} \right) W_j{}^{mn} \tn_i \eta_{\pm} \rangle = \int_{\cS}  - \frac{1}{6}e^{c\Phi} \langle \eta_{\pm} , W_{\ell_1\ell_2\ell_3}\G^{\ell_1\ell_2\ell_3} \cD \eta_{\pm} \rangle  \\
\notag
&+ \frac{q}{6} e^{c\Phi} \langle \eta_{\pm} , W_{\ell_1\ell_2\ell_3}\G^{\ell_1\ell_2\ell_3}\cA \eta_{\pm} \rangle  - \frac{1}{48} e^{c\Phi} \langle \eta_{\pm} , W_{\ell_1\ell_2\ell_3}\G^{\ell_1\ell_2\ell_3} W_{ijk}\G^{ijk} \eta_{\pm} \rangle  \\
&+ \frac{c}{12} e^{c\Phi} \langle \eta_{\pm} , \G^{i\ell_1\ell_2\ell_3} \tn_i \Phi W_{\ell_1\ell_2\ell_3} \eta_{\pm} \rangle  + \frac{1}{12} e^{c\Phi} \langle \eta_{\pm} , \G^{\ell_1\ell_2\ell_3\ell_4}\tn_{\ell_1} W_{\ell_2\ell_3\ell_4}  \eta_{\pm} \rangle \ .
\end{align}
The last term of (\ref{GGexp}) is order $\a'$, so we shall neglect it. To proceed further, we shall substitute $W_{ijk}\G^{ijk}$ in terms of $\cA$, using its definition. This produces terms proportional to the norm squared of $\cA\, \eta_{\pm}$, together with a number of counterterms. In detail, one obtains
\begin{align}
\notag
\int_{\cS}- \frac{1}{8} e^{c\Phi} \langle \eta_{\pm} , &\left( \G^{ij}\G_{mn} -  \G_{mn}\G^{ij} \right) W_j{}^{mn} \tn_i \eta_{\pm} \rangle = \int_{\cS}  - \frac{1}{6}e^{c\Phi} \langle \eta_{\pm} , W_{\ell_1\ell_2\ell_3}\G^{\ell_1\ell_2\ell_3} \cD \eta_{\pm} \rangle  \\
\notag
& + e^{c\Phi} \left(\frac{1}{48} - \frac{q}{6}\right) \langle \eta_{\pm} , \cA^{\dagger}\cA \eta_{\pm} \rangle + e^{c\Phi}\left(\frac{1}{2} - 2q\right) \langle \eta_{\pm} , \G^i \tn_i\Phi \cA \eta_{\pm} \rangle \\
\notag
& \pm e^{c\Phi} \left(\frac{1}{4}-q \right) \langle \eta_{\pm} , \G^i h_i \cA \eta_{\pm} \rangle + 3 e^{c\Phi} \langle \eta_{\pm} , \tn_i \Phi \tn^i \Phi \eta_{\pm} \rangle \pm 3e^{c\Phi}\langle\eta_{\pm} , h^i\tn_i\Phi \eta_{\pm} \rangle \\
& + \frac{3}{4} e^{c\Phi} \langle \eta_{\pm} , h_i h^i \eta_{\pm} \rangle + \frac{c}{12} e^{c\Phi} \langle \eta_{\pm} , \G^{i\ell_1\ell_2\ell_3} \tn_i \Phi W_{\ell_1\ell_2\ell_3} \eta_{\pm} \rangle + \al\ .
\end{align}
Let us focus now on the first term of (\ref{I_2 parts}). After writing $\G^{ij}$ as $\G^i\G^j - \d^{ij}$ and after integrating by parts, we have
\begin{align}
\label{I_2first}
\notag
\int_{\cS} c e^{c\Phi} \langle \eta_{\pm} , \G^{ij} \tilde{\nabla}_j \eta_{\pm} \rangle = \int_{\cS} &c e^{c\Phi} \langle \eta_{\pm} , \G^\ell \tn_\ell\Phi \G^i \tn_i \eta_{\pm} \rangle \\
&+ \frac{c}{2} e^{c\Phi} \langle \eta_{\pm} , \tn_i\tn^i \Phi  \eta_{\pm} \rangle +  \frac{c^2}{2} e^{c\Phi}\langle \eta_{\pm} , \tn_i\Phi\tn^i\Phi  \eta_{\pm} \rangle \ .
\end{align}
The first term in the RHS of (\ref{I_2first}) can be rewritten in terms of the modified Dirac operator $\cD$ after subtracting suitable terms. The second term on the RHS can be further simplified using the dilaton field equation at zeroth order in $\a'$. On performing these calculations, we have
\begin{align}
\notag
\int_{\cS} c e^{c\Phi} \langle \eta_{\pm} , \G^{ij} \tilde{\nabla}_j \eta_{\pm} \rangle = \int_{\cS} &c e^{c\Phi} \langle \eta_{\pm} , \G^\ell \tn_\ell\Phi \cD \eta_{\pm} \rangle -\frac{c}{24} e^{c\Phi} \langle \eta_{\pm} , W_{ijk}W^{ijk} \eta_{\pm} \rangle \\
\notag
& + c\left( \frac{1}{8} - q \right) e^{c\Phi} \langle \eta_{\pm} , \G^{i\ell_1\ell_2\ell_3} \tn_i\Phi W_{\ell_1\ell_2\ell_3} \eta_{\pm} \rangle + \frac{c}{4} e^{c\Phi} \langle \eta_{\pm} , h_i h^i \eta_{\pm} \rangle \\
\notag
&+ 12c \left(\frac{1}{12}+ \frac{c}{24} + q \right) e^{c\Phi} \langle \eta_{\pm} , \tn_i \Phi \tn^i \Phi \eta_{\pm} \rangle \\
&+ 6c\left(\frac{1}{12}\pm q\right)e^{c\Phi}\langle\eta_{\pm} , h^i\tn_i\Phi \eta_{\pm} \rangle + \al~.
\end{align}
Let us now focus on $\cI_3$. Recall that
\be
\label{Ricci scalar}
\G^{ij}\tn_i\tn_j \eta_{\pm} = - \frac{1}{4} \tilde{R}\, \eta_{\pm} \ .
\ee
Therefore after integrating by parts and using (\ref{Ricci scalar}) neglecting $\a'$ corrections from Einstein equations, $\cI_3$ becomes
\begin{align}
\notag
\cI_3 = \int_{\cS} - \frac{5}{48} e^{c\Phi} \langle \eta_{\pm} , W_{ijk}W^{ijk} \eta_{\pm} \rangle +& e^{c\Phi} \langle \eta_{\pm} , \tn_i\Phi \tn^i \Phi \eta_{\pm} \rangle + \frac{1}{4} e^{c\Phi} \langle \eta_{\pm} , h_i h^i \eta_{\pm} \rangle \\
& + e^{c\Phi}\langle\eta_{\pm} , h^i\tn_i\Phi \eta_{\pm} \rangle + \al \ .
\end{align}
Collecting together all terms and substituting $h_ih^i$ by inverting the zeroth order in $\a'$  dilaton filed equation, one finally gets
\begin{align}
\label{I final}
\notag
\cI = \int_{\cS} & e^{c\Phi} \langle \eta_{\pm} , \left(c\G^\ell\tn_\ell\Phi - \frac{1}{6} W_{\ell_1\ell_2\ell_3}\G^{\ell_1\ell_2\ell_3} + 2(\k - q) \cA^{\dagger} \right) \cD \eta_{\pm} \rangle \\
\notag
& + (8\k^2 -2\k q- \frac{q}{12} + q^2) e^{c\Phi} \langle \eta_{\pm} , \cA^{\dagger} \cA \eta_{\pm} \rangle \\
\notag
& + \frac{3}{4}\left( q - \frac{1}{12}\right) e^{c\Phi} \langle \eta_{\pm} , W^i{}_{\ell_1\ell_2} W^{i\ell_3\ell_4} \G^{\ell_1\ell_2\ell_3\ell_4} \eta_{\pm} \rangle \\
\notag
&-  c \left(q -\frac{1}{12}\right) e^{c\Phi} \langle \eta_{\pm} , \G^{i\ell_1\ell_2\ell_3}\tn_i\Phi W_{\ell_1\ell_2\ell_3} \eta_{\pm} \rangle \\
\notag
& + 6 \left(\frac{1}{12} + q + \frac{c}{12}\right) e^{c\Phi} \langle \eta_{\pm} , \tn_i\tn^i \Phi \eta_{\pm} \rangle \\
\notag
&+ 12c \left( q + \frac{c}{24} \right) e^{c\Phi} \langle \eta_{\pm}, \tn_i\Phi\tn^i\Phi \eta_{\pm} \rangle \\
& + \left( \frac{1}{2}-6q \pm 6q(c+2)\right) e^{c\Phi}\langle\eta_{\pm} , h^i\tn_i\Phi \eta_{\pm} \rangle + \al  \ .
\end{align}
In order to eliminate the term $\langle \eta_\pm, W^i{}_{\ell_1 \ell_2}
W_{i \ell_3 \ell_4} \Gamma^{\ell_1 \ell_2 \ell_3 \ell_4} \eta_\pm \rangle$,
which has no sign and cannot be rewritten in terms of ${\cal{D}}$
or $\cA^\dagger \cA$,
we must set
\bea
q= {1 \over 12} + \al \ .
\eea
and then in order to eliminate the $\langle \eta_\pm, \hn^i \hn_i \Phi \eta_\pm \rangle$
term we must further set
\bea
c=-2+\al \ .
\eea
Then (\ref{I final}) simplifies to
\be
\cI = \int_{\cS}  e^{-2\Phi} \langle \eta_{\pm} , \Psi \cD \eta_{\pm} \rangle + \left(8\k^2 -\frac{\k}{6}\right) \int_{\cS} e^{-2\Phi} \parallel \cA \, \eta_{\pm} \parallel^2 +\, \al \ ,
\ee
where
\be
\label{Psi}
\Psi \equiv -2\G^\ell\tn_\ell\Phi - \frac{1}{6} W_{\ell_1\ell_2\ell_3}\G^{\ell_1\ell_2\ell_3} + 2\left(\k - \frac{1}{12}\right) \cA^{\dagger} \ .
\ee

\subsection{Computations at first order in $\a'$}
In this section we shall consider corrections at first order in $\a'$. $\cI_2$ and $\cI_3$ gain $\a'$ corrections from bosonic field equations and Bianchi identity, while $\cI_1$ does not. Therefore we have
\begin{align}
\notag
\cI_1 = &\int_{\cS} e^{c\Phi} 2(\k- q) \langle \eta_{\pm} , \cA^{\dagger} \cD \eta_{\pm} \rangle + e^{c\Phi} (8\k^2 - 2 \k q + q^2) \langle \eta_{\pm} , \cA^{\dagger}\cA \eta_{\pm} \rangle   \\
&- \frac{1}{8} e^{c\Phi}\langle \eta_{\pm} ,  W^i{}_{\ell_1\ell_2}W_{i\ell_3\ell_4}\G^{\ell_1\ell_2\ell_3\ell_4}\eta_{\pm} \rangle  + \frac{1}{16}e^{c\Phi} \langle\eta_{\pm} , W_{ijk}W^{ijk}\eta_{\pm} \rangle + \alsq\ ,
\end{align}
and
\begin{align}
\notag
\cI_2 = \int_{\cS}  &\ c e^{c\Phi} \langle \eta_{\pm} , \left(\G^\ell \tn_\ell\Phi  -\frac{1}{6}W_{\ell_1\ell_2\ell_3}\G^{\ell_1\ell_2\ell_3}\right)\cD \eta_{\pm} \rangle -\frac{c}{24} e^{c\Phi} \langle \eta_{\pm} , W_{ijk}W^{ijk} \eta_{\pm} \rangle \\
\notag
& + c\left( \frac{5}{24} - q \right) e^{c\Phi} \langle \eta_{\pm} , \G^{i\ell_1\ell_2\ell_3} \tn_i\Phi W_{\ell_1\ell_2\ell_3} \eta_{\pm} \rangle + e^{c\Phi} \left(\frac{1}{48} - \frac{q}{6}\right) \langle \eta_{\pm} , \cA^{\dagger}\cA \eta_{\pm} \rangle \\
\notag
&  + e^{c\Phi}\left(\frac{1}{2} - 2q\right) \langle \eta_{\pm} , \G^i \tn_i\Phi \cA \eta_{\pm} \rangle \pm e^{c\Phi} \left(\frac{1}{4}-q \right) \langle \eta_{\pm} , \G^i h_i \cA \eta_{\pm} \rangle \\
\notag
& + \left(\frac{3}{4} + \frac{c}{4}\right) e^{c\Phi} \langle \eta_{\pm} , h_i h^i \eta_{\pm} \rangle + \left(c + \frac{c^2}{2} + 12cq + 3 \right) e^{c\Phi} \langle \eta_{\pm} , \tn_i \Phi \tn^i \Phi \eta_{\pm} \rangle  \\
\notag
& + \left(\frac{c}{2}\pm 6cq\pm 3\right) e^{c\Phi}\langle\eta_{\pm} , h^i\tn_i\Phi \eta_{\pm} \rangle \\
\notag
&-\frac{1}{24}e^{c\Phi} \langle \eta_{\pm} , \G^{\ell_1\ell_2\ell_3\ell_4} \tn_{\ell_1}W_{\ell_2\ell_3\ell_4} \eta_{\pm} \rangle + \a' \frac{c}{32} e^{c\Phi} \bigg( -2 \langle \eta_{\pm},  dh_{ij}dh^{ij} \eta_{\pm} \rangle  \\
&+ \langle \eta_{\pm} , \check{\tilde{R}}_{\ell_1\ell_2,\ell_3\ell_4}\check{\tilde{R}}^{\ell_1\ell_2,\ell_3\ell_4} \eta_{\pm} \rangle - \langle \eta_{\pm} , \tilde{F}_{ij}{}^{ab}\tilde{F}^{ij}{}_{ab} \eta_{\pm} \rangle \bigg) + \alsq \ ,
\end{align}
and
\begin{align}
\notag
\cI_3 = \int_{\cS} & - \frac{5}{48} e^{c\Phi} \langle \eta_{\pm} , W_{ijk}W^{ijk} \eta_{\pm} \rangle + e^{c\Phi} \langle \eta_{\pm} , \tn_i\Phi \tn^i \Phi \eta_{\pm} \rangle + \frac{1}{4} e^{c\Phi} \langle \eta_{\pm} , h_i h^i \eta_{\pm} \rangle  \\
\notag
&+ e^{c\Phi}\langle\eta_{\pm} , h^i\tn_i\Phi \eta_{\pm} \rangle +\a' \frac{3}{32} e^{c\Phi} \bigg( -2 \langle \eta_{\pm},  dh_{ij}dh^{ij} \eta_{\pm} \rangle \\
&+ \langle \eta_{\pm} , \check{\tilde{R}}_{\ell_1\ell_2,\ell_3\ell_4}\check{\tilde{R}}^{\ell_1\ell_2,\ell_3\ell_4} \eta_{\pm} \rangle - \langle \eta_{\pm} , \tilde{F}_{ij}{}^{ab}\tilde{F}^{ij}{}_{ab} \eta_{\pm} \rangle \bigg) + \alsq \ .
\end{align}
Combining all together and considering $\a'$ corrections from substituting $h_ih^i$ by inverting the dilaton field equations, we have
\begin{align}
\label{I alpha' 1}
\notag
\cI = \int_{\cS} & e^{c\Phi} \langle \eta_{\pm} , \left(c\G^\ell\tn_\ell\Phi - \frac{1}{6} W_{\ell_1\ell_2\ell_3}\G^{\ell_1\ell_2\ell_3} + 2(\k - q) \cA^{\dagger} \right) \cD \eta_{\pm} \rangle \\
\notag
& + (8\k^2 -2\k q- \frac{q}{12} + q^2) e^{c\Phi} \langle \eta_{\pm} , \cA^{\dagger} \cA \eta_{\pm} \rangle \\
\notag
& + \frac{3}{4}\left( q - \frac{1}{12}\right) e^{c\Phi} \langle \eta_{\pm} , W^i{}_{\ell_1\ell_2} W^{i\ell_3\ell_4} \G^{\ell_1\ell_2\ell_3\ell_4} \eta_{\pm} \rangle \\
\notag
&-  c \left(q -\frac{1}{12}\right) e^{c\Phi} \langle \eta_{\pm} , \G^{i\ell_1\ell_2\ell_3}\tn_i\Phi W_{\ell_1\ell_2\ell_3} \eta_{\pm} \rangle + 12c \left( q + \frac{c}{24} \right) e^{c\Phi} \langle \eta_{\pm}, \tn_i\Phi\tn^i\Phi \eta_{\pm} \rangle \\
\notag
& + 6 \left(\frac{1}{12} + q + \frac{c}{12}\right) e^{c\Phi} \langle \eta_{\pm} , \tn_i\tn^i \Phi \eta_{\pm} \rangle  + \left( \frac{1}{2}-6q \pm 6q(c+2)\right)e^{c\Phi}\langle\eta_{\pm} , h^i\tn_i\Phi \eta_{\pm} \rangle \\
\notag
&+ \frac{\a'}{64}e^{c\Phi} \bigg( 2 \langle\eta_{\pm}, \G^{\ell_1\ell_2\ell_3\ell_4}dh_{\ell_1\ell_2}dh_{\ell_3\ell_4}\rangle - \langle\eta_{\pm}, \G^{\ell_1\ell_2\ell_3\ell_4}\check{\tilde{R}}_{\ell_1\ell_2, ij}\check{\tilde{R}}_{\ell_3\ell_4,}{}^{ij}\eta_{\pm}\rangle  \\
\notag
& \hspace{7cm}+ \langle\eta_{\pm}, \G^{\ell_1\ell_2\ell_3\ell_4}\tilde{F}_{\ell_1\ell_2,\, ab}\tilde{F}_{\ell_3\ell_4}{}^{ab}\eta_{\pm}\rangle \bigg)  \\
\notag
&+ \a' \frac{3}{8}\left(\frac{1}{6}-q\right) e^{c\Phi}\bigg( -2 \langle \eta_{\pm},  dh_{ij}dh^{ij} \eta_{\pm} \rangle \\
&+ \langle \eta_{\pm} , \check{\tilde{R}}_{\ell_1\ell_2,\ell_3\ell_4}\check{\tilde{R}}^{\ell_1\ell_2,\ell_3\ell_4} \eta_{\pm} \rangle - \langle \eta_{\pm} , \tilde{F}_{ij}{}^{ab}\tilde{F}^{ij}{}_{ab} \eta_{\pm} \rangle \bigg) + \alsq \ .
\end{align}

To further simplify (\ref{I alpha' 1}), we note the following identity
\be
\label{dh identity}
\langle \eta_{\pm} , \G^{\ell_1\ell_2\ell_3\ell_4} dh_{\ell_1\ell_2} dh_{\ell_3\ell_4} \eta_{\pm} \rangle = \langle \eta_{\pm} , \G^{\ell_1\ell_2}dh_{\ell_1\ell_2} \G^{\ell_3\ell_4}dh_{\ell_3\ell_4} \eta_{\pm} \rangle + 2 \langle \eta_{\pm} , dh_{ij}dh^{ij} \eta_{\pm} \rangle  \ .
\ee
Identities analogous to (\ref{dh identity}) hold also for the terms which involve $\check{\tilde{R}}_{ij,k\ell}$ and $\tilde{F}_{ij}{}^{ab}$.
This leads to
\begin{align}
\label{I alpha' 2}
\notag
\cI = \int_{\cS} & e^{c\Phi} \langle \eta_{\pm} , \left(c\G^\ell\tn_\ell\Phi - \frac{1}{6} W_{\ell_1\ell_2\ell_3}\G^{\ell_1\ell_2\ell_3} + 2(\k - q) \cA^{\dagger} \right) \cD \eta_{\pm} \rangle \\
\notag
& + (8\k^2 -2\k q- \frac{q}{12} + q^2) e^{c\Phi} \langle \eta_{\pm} , \cA^{\dagger} \cA \eta_{\pm} \rangle \\
\notag
& + \frac{3}{4}\left( q - \frac{1}{12}\right) e^{c\Phi} \langle \eta_{\pm} , W^i{}_{\ell_1\ell_2} W^{i\ell_3\ell_4} \G^{\ell_1\ell_2\ell_3\ell_4} \eta_{\pm} \rangle \\
\notag
&-  c \left(q -\frac{1}{12}\right) e^{c\Phi} \langle \eta_{\pm} , \G^{i\ell_1\ell_2\ell_3}\tn_i\Phi W_{\ell_1\ell_2\ell_3} \eta_{\pm} \rangle + 12c \left( q + \frac{c}{24} \right) e^{c\Phi} \langle \eta_{\pm}, \tn_i\Phi\tn^i\Phi \eta_{\pm} \rangle \\
\notag
& + 6 \left(\frac{1}{12} + q + \frac{c}{12}\right) e^{c\Phi} \langle \eta_{\pm} , \tn_i\tn^i \Phi \eta_{\pm} \rangle  + \left( \frac{1}{2}-6q \pm 6q(c+2)\right)e^{c\Phi}\langle\eta_{\pm} , h^i\tn_i\Phi \eta_{\pm} \rangle\\
\notag
&+{3 \over 8}\alpha' (q-{1 \over 12})e^{c \Phi}
\bigg(2 dh_{ij} dh^{ij} + {\tilde{F}}_{ij}{}^{ab} {\tilde{F}}^{ij}{}_{ab}
- \check{\tilde{R}}_{\ell_1 \ell_2,\ell_3 \ell_4}
\check{\tilde{R}}^{ \ell_1 \ell_2, \ell_3 \ell_4} \bigg) \parallel \eta_{\pm} \parallel^2 \\
\notag
& - \frac{\a'}{32} e^{c\Phi} \parallel \slashed{dh}\, \eta_{\pm}\parallel ^2
- \frac{\a'}{64} e^{c\Phi} \parallel \slashed{\tilde{F}}\, \eta_{\pm} \parallel ^2 + \frac{\a'}{64} e^{c\Phi}\langle \check{\tilde{R}}_{\ell_1\ell_2,\, ij}\G^{\ell_1\ell_2}\eta_{\pm}, \check{\tilde{R}}^{ ij}_{\ell_3\ell_4,}\G^{\ell_3\ell_4}\eta_{\pm}\rangle
\\ \notag
 &+ \, \alsq \ .
\end{align}
In order to eliminate the term $\langle \eta_\pm, W^i{}_{\ell_1 \ell_2}
W_{i \ell_3 \ell_4} \Gamma^{\ell_1 \ell_2 \ell_3 \ell_4} \eta_\pm \rangle$,
which has no sign and cannot be rewritten in terms of ${\cal{D}}$
or $\cA^\dagger \cA$,
we must set
\bea
q= {1 \over 12} + \alsq \ .
\eea
and then in order to eliminate the $\langle \eta_\pm, \hn^i \hn_i \Phi \eta_\pm \rangle$
term we must further set
\bea
c=-2+\alsq \ .
\eea
Then (\ref{I alpha' 2}) is significantly simplified to
\begin{align}
\notag
\cI  = &\left(8\k^2 - \frac{1}{6} \k \right) \int_{\cS} e^{-2 \Phi} \parallel \cA\,  \eta_{\pm} \parallel^2
+ \int_{\cS} e^{-2\Phi} \langle \eta_{\pm}, \Psi \cD \eta_{\pm} \rangle \\
&- \frac{\a'}{64} \int_{\cS} e^{-2\Phi} \left( 2 \parallel \slashed{dh}\, \eta_{\pm} \parallel^2 + \parallel \slashed{\tilde{F}} \eta_{\pm} \parallel^2 - \langle \check{\tilde{R}}_{\ell_1\ell_2,\, ij}\G^{\ell_1\ell_2}\eta_{\pm}, \check{\tilde{R}}^{ij}_{\ell_3\ell_4,}\G^{\ell_3\ell_4}\eta_{\pm}\rangle \right) + \alsq\ ,
\end{align}
where $\Psi$ is defined in (\ref{Psi}).

\appendix{$AdS_{n+1}$ as warped product over $AdS_n$}
The $AdS_{n+1}$ space can be written  as a warped product over $AdS_n$. This has been observed before in \cite{strominger} for $AdS_3$ and elsewhere, see eg \cite{slicex}. For this, we
label all geometrical objects defined on $AdS_{n+1}$ and $AdS_n$ by $n+1$ and $n$ respectively, e.g. $ds^2_{n+1}$ is the metric on $AdS_{n+1}$ and $ds^2_n$ is the metric on $AdS_n$. In principle $AdS_{n+1}$ and $AdS_n$ can have different radii, which are indicated by $\ell_{n+1}$ and $\ell_n$ respectively.
Coordinates on $AdS_{n+1}$ are taken to be as follows
\be
x^{I} = (x^0, x^i) \ , \qquad\qquad x^0 \equiv y \ , \qquad i= 1, ... , n \ .
\ee

We shall begin with an Ans{\"a}tz for the metric on $AdS_{n+1}$ as a warped product over $AdS_n$, i.e.
\be
\label{metric ansatz n+1}
ds_{n+1}^2 = dy^2 + f(y)^2 ds_n^2 \ .
\ee
We want to determine the necessary and sufficient conditions to impose on $f(y)$ in order for $ds^2_{n+1}$ to be the metric on $AdS_{n+1}$. To succeed, we have to impose the fact $AdS_{n+1}$ is a maximally symmetric space. Locally, the necessary and sufficient condition is that the Riemann tensor must assume the following form
\be
\label{maximally_symm}
R^{(n+1)}_{IJKL} = -\frac{1}{\ell^2_{n+1}} \left(g^{(n+1)}_{IK}g^{(n+1)}_{JL} - g^{(n+1)}_{JK}g^{(n+1)}_{IL} \right) \ ,
\ee
Equation (\ref{maximally_symm}) implies also that the metric (\ref{metric ansatz n+1}) is Einstein and the curvature scalar is constant and negative, i.e.
\be
\label{Einstein}
R^{(n+1)}_{IJ} = - \frac{n}{\ell^2_{n+1}} g^{(n+1)}_{IJ} \ , \qquad\qquad R^{(n+1)}= -\frac{1}{\ell^2_{n+1}}n(n+1) \ .
\ee

The non-vanishing Christoffel symbols of (\ref{metric ansatz n+1}) are:
\be
\G^{(n+1)\, k}_{\qquad i \  0} = \frac{f'(y)}{f(y)} \d^k{}_i \ , \qquad \G^{(n+1)\, 0}_{\qquad i \  j} = - f(y)f'(y)g^{(n+1)}_{ij} \ , \qquad
\G^{(n+1)\, k}_{\qquad i \  j} = \G^{(n)\, k}_{\quad\, i \  j}  \ .
\ee
The non-vanishing Riemann tensor components are
\begin{align}
\notag
R^{(n+1)}{}_{i0,}{}^{k}{}_0 &= - \frac{f''(y)}{f(y)} \d^k{}_i \ , \\
\notag
R^{(n+1)}{}_{i0,}{}^0{}_{\ell} &= f(y)f''(y) g^{(n)}_{i\ell} \ , \\
R^{(n+1)}{}_{ij,}{}^k{}_{\ell} & =  R^{(n)}{}_{ij,}{}^k{}_{\ell} + f'(y)^2\left(\d^k{}_j g^{(n)}_{i\ell} - \d^k{}_i g^{(n)}_{j\ell} \right) \ ,
\end{align}
and
\begin{align}
\notag
R^{(n+1)}_{i0,k0} &= - f(y)f''(y) g_{ik}^{(n)} \ , \\
R^{(n+1)}_{ij, kl} &= f(y)^2 R^{(n)}_{ij,kl} - f(y)^2f'(y)^2\left(g^{(n)}_{ik}g^{(n)}_{jl} - g^{(n)}_{jk}g^{(n)}_{il} \right) \ .
\end{align}
The non-vanishing Ricci tensor components are
\begin{align}
\label{Ricci n+1}
\notag
R^{(n+1)}_{00} &= - n \frac{f''(y)}{f(y)} \ , \\
R^{(n+1)}_{ij} &= R^{(n)}_{ij} + \left[f'(y)^2(1-n) - f(y)f''(y)\right] g^{(n)}_{ij} \ .
\end{align}
The Riemann tensor on $AdS_n$ must assume the following form
\be
R^{(n)}_{ijk\ell} = - \frac{1}{\ell^2_{n}}\left(g_{ik}^{(n)}g_{j\ell}^{(n)} - g_{jk}^{(n)}g_{i\ell}^{(n)} \right) \ .
\ee
Now we impose (\ref{maximally_symm}). The $(i0,k0)$-components provide the first ordinary differential equation for $f$
\be
\label{1_ODE}
f''(y) = \frac{1}{\ell^2_{n+1}} f(y) \ .
\ee
The $(ij,kl)$-components provide the second ordinary differential equation for $f$
\be
\label{2_ODE}
f'(y)^2 - \frac{1}{\ell^2_{n+1}} f(y)^2 + \frac{1}{\ell^2_n} = 0 \ .
\ee
Since equations in (\ref{Einstein}) are derived from (\ref{maximally_symm}), they would imply again (\ref{1_ODE}) and (\ref{2_ODE}), so there is nothing further to be learned from those conditions.
The general solution of (\ref{1_ODE}) and (\ref{2_ODE}) is
\be
\label{warped}
f(y) = \alpha \cosh\left(\frac{y}{\ell_{n+1}}\right) + \beta \sinh\left(\frac{y}{\ell_{n+1}}\right)\ ,
\ee
where $\alpha$ and $\beta$ are constants which satisfy
\be
\alpha^2 - \beta^2 = \frac{\ell^2_{n+1}}{\ell^2_{n}} \ .
\ee
The solution (\ref{warped}) leads us to the following conclusions
\begin{enumerate}
\item if $y\in (-\infty, +\infty)$, then locally the $AdS_3$ metric can be written as $AdS_2\times_w \mathbb{R}$.
\item if $y\in [0, 1]$, then locally the $AdS_3$ metric can be written as  is $AdS_2\times_w [0, 1]$ as the warp factor is not periodic.
\item if $y\in [0, 1]$ and force periodicity on $y$, then the metric of $AdS_2\times_w S^1$ is discontinuous as the warp factor is not periodic.
\end{enumerate}
From the perspective of near horizon geometries, the first case violates the compactness condition of the partial horizon section. The second case implies that the spatial horizon
 section has a  boundary. The third case violates  smoothness condition since (\ref{warped}) is not periodic.
Hence  all cases violate one or more  of the  assumptions required to prove that there are no  $AdS_2$ horizons in the heterotic theory.

\end{document}